\documentclass[journal]{IEEEtran}
\usepackage{amsmath,amsfonts}

\usepackage{array}
\usepackage{textcomp}
\usepackage{stfloats}
\usepackage{url}
\usepackage{verbatim}
\usepackage{graphicx}
\usepackage{cite}
\usepackage{multirow}
\usepackage{booktabs}

\usepackage{color}
\usepackage{hyperref}
\usepackage{graphicx}
\usepackage{amsmath}
\usepackage{amssymb}
\usepackage{cleveref}
\usepackage[section]{placeins}
\usepackage{float}
\usepackage{diagbox}

\usepackage{epstopdf}
\usepackage{amsmath}
\usepackage{amssymb}
\usepackage{multirow}
\usepackage{hyperref}

\usepackage{xcolor}

\usepackage{graphicx}%
\usepackage{multirow}%
\usepackage{amsmath,amssymb,amsfonts}%
\usepackage{amsthm}%
\usepackage{mathrsfs}%
\usepackage{xcolor}%
\usepackage{textcomp}%
\usepackage{manyfoot}%
\usepackage{booktabs}%
\usepackage{algorithm}%
\usepackage{algorithmicx}%
\usepackage{algpseudocode}%
\usepackage{listings}%

\usepackage{array}
\usepackage[caption=false,font=scriptsize,labelfont=sf,textfont=sf]{subfig}
\usepackage{stfloats}
\usepackage{url}
\usepackage{verbatim}
\usepackage{color}
\usepackage{hyperref}
\usepackage{cleveref}
\usepackage[section]{placeins}
\usepackage{float}
\usepackage{diagbox}
\usepackage{epstopdf}
\usepackage [utf8]{inputenc}
\usepackage{bbding}
\usepackage{booktabs}
\usepackage{makecell}
\usepackage{tabularx}
\usepackage{tikz}
\usepackage{ltxtable}
\usepackage{wrapfig}
\usetikzlibrary{shapes.geometric, arrows}

\usepackage{hyperref}

\hypersetup{
    colorlinks=true, 
    urlcolor=magenta,   
}

\usepackage{setspace}

\usepackage{url}

\ifCLASSINFOpdf
\else
\fi
\begin{document}

%
\title{Mean Masked Autoencoder with Flow-Mixing for Encrypted Traffic Classification}
%
%
%
%

\author{Xiao Liu, Xiaowei Fu, Fuxiang Huang, and Lei Zhang,~\IEEEmembership{Senior Member,~IEEE}
\thanks{This work was partially supported by National Natural Science Fund of China under Grants 92570110 and 62271090, Chongqing Natural Science Fund under Grant CSTB2024NSCQ-JQX0038, and National Youth Talent Project. \textit{(Corresponding author: Lei Zhang)}}
\thanks{\IEEEcompsocthanksitem X. Liu, X. Fu and L. Zhang are with the School of Microelectronics and Communication Engineering, Chongqing University, Chongqing 400044, China.
(E-mail: liuxiao@stu.cqu.edu.cn, xwfu@cqu.edu.cn, leizhang@cqu.edu.cn,)

Fuxiang Huang is with the School of Data Science, Lingnan University, Hong Kong, China.
(E-mail: fxhuang1995@gmail.com)
}
\thanks{Manuscript received April 19, 2021; revised August 16, 2021.}}

\markboth{Journal of \LaTeX\ Class Files,~Vol.~14, No.~8, August~2021}%
{Shell \MakeLowercase{\textit{et al.}}: Bare Demo of IEEEtran.cls for Computer Society Journals}
%



\IEEEtitleabstractindextext{%
\begin{abstract}
Network traffic classification using self-supervised pre-training models based on Masked Autoencoders (MAE) has demonstrated a huge potential. However, existing methods are confined to isolated byte-level reconstruction of individual flows, lacking adequate perception of the multi-granularity contextual relationship in traffic. To address this limitation, we propose Mean MAE (MMAE), a teacher-student MAE paradigm with flow mixing strategy for building encrypted traffic pre-training model. MMAE employs a self-distillation mechanism for teacher-student interaction, where the teacher provides unmasked flow-level semantic supervision to advance the student from local byte reconstruction to multi-granularity comprehension. To break the information bottleneck in individual flows, we introduce a dynamic Flow Mixing (FlowMix) strategy to replace traditional random masking mechanism. By constructing challenging cross-flow mixed samples with interferences, it compels the model to learn discriminative representations from distorted tokens. Furthermore, we design a Packet-importance aware Mask Predictor (PMP) equipped with an attention bias mechanism that leverages packet-level side-channel statistics to dynamically mask tokens with high semantic density. Numerous experiments on a number of datasets covering encrypted applications, malware, and attack traffic demonstrate that MMAE achieves state-of-the-art performance. The code is available at \href{https://github.com/lx6c78/MMAE}{https://github.com/lx6c78/MMAE}

\end{abstract}

\begin{IEEEkeywords}
Encrypted Traffic Classification, Masked Autoencoder, Pre-training,  Self-Distillation.
\end{IEEEkeywords}}

\maketitle

\IEEEdisplaynontitleabstractindextext

%
\IEEEpeerreviewmaketitle

\section{Introduction}\label{sec1}
\IEEEPARstart{W}{ith} the rapid development of the Internet, network traffic classification has become a key challenge. 
Traffic data streams provide critical insights into network behavior through payloads and metadata. Effective traffic monitoring and analysis are crucial for network management and security, ensuring Quality of Service (QoS) and detecting malicious activities \cite{network_survey}, \cite{PERT}. Consequently, network traffic classification has emerged as a vital research area. The main objective is to identify potential threats, various applications and services. However, the encryption protocols (e.g., TLS) and anonymity networks (e.g., VPNs, Tor) imposes further challenge. 


\begin{figure}[t]
\begin{center}
\includegraphics[width=1\linewidth]{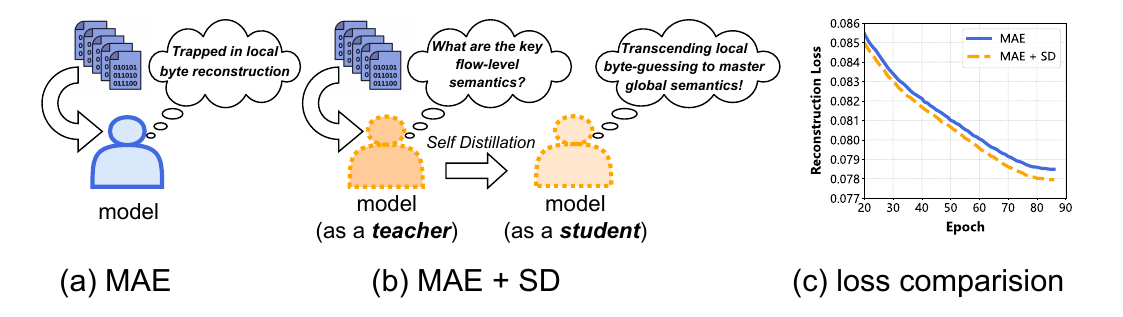}
\end{center}
\vspace{-0.4cm}
   \caption{Comparison of reconstruction loss during pre-training between MAE and its variant with Self-Distillation strategy (i.e., MAE+SD). By introducing flow-level semantics via SD, the loss is consistently lower.}
\label{loss_comp}
\end{figure}


Recently, pre-training methods \cite{Et_bert}, \cite{YATC}, \cite{Netmamba}, \cite{Trafficformer}, \cite{FlowletFormer} towards network traffic classification have shown immense promise and superior performance. Based on learned generalizable representations from massive unlabeled data, these models can be effectively fine-tuned on limited labeled datasets. These pre-training methods typically employ a strategy of \textit{mask-then-reconstruct}. For instance, ET-BERT \cite{Et_bert}, FlowletFormer \cite{FlowletFormer} and Trafficformer \cite{Trafficformer} adopt masking strategies derived from language modeling. YaTC \cite{YATC} treats each traffic flow as an image to perform masked flow reconstruction, whereas NetMamba \cite{Netmamba} employs a 1D sequence-specific masking strategy for pre-training objectives.

Despite the above progress, such Masked Autoencoder (MAE) pre-training paradigms for network traffic classification still face critical challenges. These limitations can be summarized in two aspects.
1) \textit{Current MAE methods are restricted to isolated byte-level reconstruction of individual flows}, failing to capture the inherent multi-granularity structure of network traffic from packet or flow.
2) \textit{The random masking strategy in MAE focuses only on byte-level but ignores traffic-specific attributes (e.g., packet-level and flow-level)}, struggling to construct sufficiently challenging pretext tasks and hindering the model from learning discriminative semantics. The rational behind the first limitation is that accurate classification requires moving beyond the perception of local regions within a single flow \cite{zhao2024novel}, \cite{sirinam2018deep}. For example, different applications often contain similar local traffic patterns, which means that relying solely on local byte-level information can only provide limited application-specific semantics and low-confidence classification results \cite{chen2023classify}, \cite{wang2025mfsi}. The rational behind the second point is that since not all tokens possess equal information density, random masking often leads to targeting less informative tokens for reconstruction. Such trivial pretext tasks hinder learning useful representations. Therefore, \textit{exploring, harnessing and improving MAE is challenging but critical for encrypted traffic classification}. 

\begin{figure}[t]
\begin{center}
\includegraphics[width=1\linewidth]{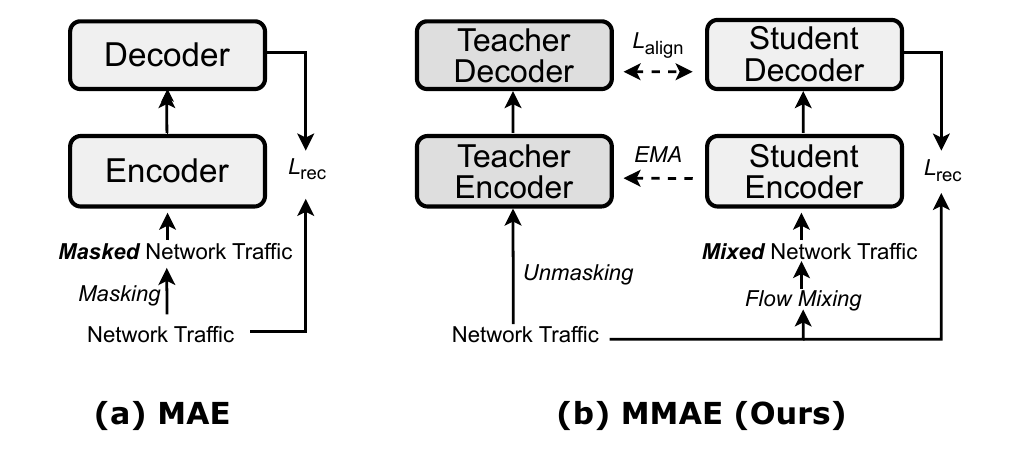}
\end{center}
   \caption{Comparison of pre-training paradigms. (a) Standard MAE with a random masking strategy in an isolated single flow.
(b) MMAE (ours): A novel teacher-student pre-training architecture with a flow-mixing strategy, where the teacher is a copy of student and interacted via self-distillation strategy under the unmasked flow-level semantic supervision from the teacher. 
}
\label{framework_comp}
\end{figure}


To validate the above suspicions that the conventional MAE inadequately captures traffic information, we conduct a pilot experiment. Specifically, we deploy a self-distillation (SD) strategy into the MAE framework. In this setup, a copy of MAE without masking is deployed as a teacher to extract flow-level global semantics to provide richer supervision, while the student model still performs standard masked token reconstruction but under an extra supervision from the teacher. As shown in Fig.~\ref{loss_comp}, we find that the reconstruction loss of the student model continues to decrease by deploying the SD strategy. This indicates that standard MAE pre-training strategy fails to fully explore useful information of traffic. 

To better represent traffic in multi-granularity, we rethink the pre-training paradigm and propose a Mean Masked Autoencoder (MMAE) with flow mixing mechanism. This framework extracts rich representations from unlabeled traffic data, achieving a comprehensive understanding of multi-granularity semantic features. Specifically, MMAE consists of a MAE (i.e., student) and its copy (i.e., teacher), which interacts via a self-distillation strategy based on EMA (Exponential Moving Average). This is the basic connotation of \textit{Mean} MAE. 
Fig. \ref{framework_comp} illustrates the key distinctions between MAE and our MMAE. Conventional MAE is limited to single-flow reconstruction with random masking mechanism, which tends to local byte guessing, failing to capture inter-flow class boundaries and discriminative semantics. To break the single-flow information bottleneck, MMAE allows the teacher (a copy of MAE without masking) to extract complete semantics from the unmasked input and provide flow-level supervision. 

Distinct encrypted traffic often exhibit similarities when viewed as isolated byte fragments. But in fact, their differences are reflected through multi-granularity contexts. Consequently, conventional random masking strategy of MAE in only byte-level is passive and trivial.
To construct more challenging traffic-aware pretext tasks beneficial to discriminative semantic learning, we propose dynamic flow mixing (FlowMix).  \textit{The core idea is to inject cross-flow interference and replace the conventional random masking strategy}. To ensure this interference is challenging rather than trivially distinguishable, in FlowMix, we introduce a Statistics-based Flow Matcher (SFM) leveraging statistical priors to pair physically similar flows. After selecting the paired flows through SFM, we deploy a Packet-importance aware Mask Predictor (PMP) to determine where to inject interference. 
Specifically, PMP dynamically identifies regions of high semantic density, and then perform Dynamic Mixed Masking (DMM) to replace these critical regions with the matched distorted tokens. Thus the student is compelling to learn highly discriminative representations.

\begin{figure}[t]
\begin{center}
\includegraphics[width=0.9 \linewidth]{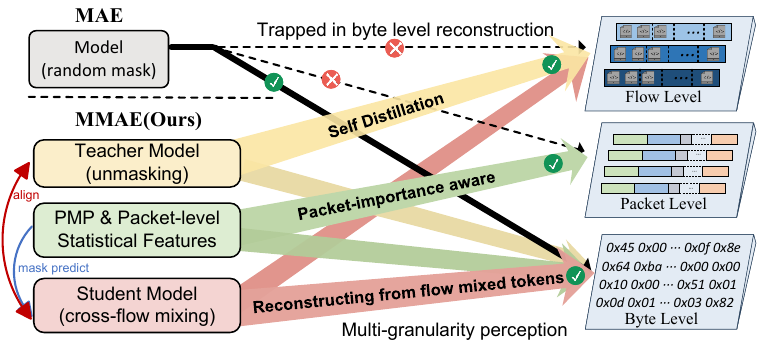}
\end{center}
   \caption{Comparison of hierarchical semantic extraction capabilities between MAE and the proposed MMAE. MAE relies solely on a single flow and random masking for reconstruction. Our MMAE incorporates cross-flow mixing and packet-level side-channel statistical features. Within the self-distillation teacher-student architecture, flow-level reference features are used as supervision for reconstruction and mask prediction tasks.}
\label{two_mask_strategy}
\vspace{-0.3cm}
\end{figure}

This above design inherently guarantees multi-granularity perception. 1) \textit{Byte-level} details are captured by reconstructing from the challenging flow mixed tokens. 2) \textit{Packet-level} dependencies are learned through a Packet-importance aware Mask Predictor (PMP), which is guided by packet-level physical priors and subsequently fed back to dynamically construct mixed inputs for the next training iteration. 3) \textit{Flow-level} global semantics are established by aligning the student's representation with the teacher. Fig. \ref{two_mask_strategy} indicates
the key differences in hierarchical representational capacity, in which MAE only captures byte-level information, while \textit{MMAE covers byte-level, packet-level as well as flow-level details}.

The main contributions are summarized as follows:
\begin{itemize}
    \item
    We propose Mean Masked Autoencoder (\textbf{MMAE}), a novel self-supervised pre-training paradigm for encrypted traffic classification. The framework leverages a twin architecture (teacher-student) of MAE, interacted via self-distillation under the supervision of unmasked flow-level semantics from teacher to enable the student learning multi-granular representations.
    
    \item
    We propose a dynamic Flow Mixing (\textbf{FlowMix}) strategy beyond the conventional random masking in MAE to fully explore byte-level, packet-level and flow-level information. 
    The strategy actively injects cross-flow interference to construct highly challenging pretext tasks, enabling the model to effectively delineate true traffic boundaries through multi-granularity contexts and learn discriminative semantics.

    \item
    We further design a Statistics-based Flow Matcher (SFM) and a Packet-importance aware Mask Predictor (PMP) to implement cross-flow interference in \textbf{FlowMix}. SFM leverages statistical priors to pair physically similar flows. PMP identifies semantically dense regions that are subsequently replaced via Dynamic Mixed Masking (DMM).
    
\end{itemize}


\section{Related Work}
\subsection{Traditional Methods}
Driven by the increasing complexity of network environments and the escalating demands of network management, traffic classification methods have undergone significant evolution over the past decade. Earlier approaches primarily relied on port numbers and simple rule matching \cite{snort}, \cite{zuev2005traffic}. However, the widespread adoption of encryption and traffic obfuscation techniques has rendered these plaintext- and rule-based methods largely ineffective.

Early machine learning approaches \cite{moore2005internet}, \cite{bernaille2006traffic}, \cite{panchenko2016website}, \cite{al2016adaptive}, \cite{taylor2017robust} explored classifiers such as decision trees, random forests, and SVMs for traffic classification. These techniques primarily utilized expert-designed, flow-level statistical summaries and protocol-specific features as model inputs. Although lightweight, fast, and interpretable, their performance heavily depends on time-consuming manual feature engineering, which limits their adaptability to new protocols. Consequently, such approaches struggle to flexibly adapt rapidly evolving network protocols. Despite these limitations, ML-based methods retain significant value in modern contexts. Handcrafted feature, such as traffic statistics, are relatively stable and frequently utilized to complement other methods in traffic classification \cite{umair2021efficient}, \cite{li2013hybrid}, \cite{zhang2025optimized}.

Deep learning (DL) models (e.g., CNNs, RNNs, and GNNs) have significantly improved the performance by automatically extracting representations from raw packets, such as raw bytes or key packet attributes \cite{sirinam2018deep}, \cite{shen2021accurate}, \cite{schuster2017beauty}, \cite{zhang2020autonomous}, \cite{Fs_net}.
However, these conventional neural architectures exhibit limitations in capturing long-range sequential dependencies, which are prone to inductive biases and suffer from low computational efficiency. Transformer-based methods \cite{NetST}, \cite{miett}, \cite{liu2025transeca} offer powerful sequence modeling capabilities compared to CNNs by leveraging self-attention. However, their high performance typically relies on large-scale labeled datasets, which are costly to obtain in practical applications.

\subsection{Pre-training based Methods}
Pre-training methods significantly reduce the demand of labeled training data. This paradigm typically consists of two distinct phases: \textit{pre-training stage} based on unlabeled traffic data via self-supervised learning and \textit{fine-tuning stage} based on a few labeled downstream traffic data. 
With the success of large models such as BERT\cite{bert} and GPT \cite{gpt}, pre-training methodologies have rapidly emerged as the dominant approach for sequence modeling.
Recent studies focus on Transformer-based pre-training paradigms for network traffic classification. For example, PERT \cite{PERT} directly feeds raw packet sequences into a Transformer for feature extraction. Conversely, YaTC \cite{YATC} transforms traffic into two-dimensional images and trains a visual masked autoencoder. Additionally, frameworks such as ET-BERT \cite{Et_bert} and TrafficFormer \cite{Trafficformer} partition traffic into bursts by analyzing packet transmission directions. These bursts are subsequently serialized into hexadecimal strings, and the subword tokenization techniques are applied to construct a fixed-size vocabulary.
Although these methods have achieved notable success, directly adopting local reconstruction strategies from natural language processing or computer vision often breaks the inherent properties of network traffic. Consequently, such designs fail to adequately capture the unique semantic characteristics of network traffic.

\subsection{Knowledge Distillation}
Knowledge Distillation (KD) is a classic knowledge transfer and model compression technique \cite{hinton2015distilling}. The core objective is to train a compact student model to approximate the targets generated by a complex teacher model, which enables the model to inherit the strong generalization capabilities.
Self-distillation deploys the same underlying architecture for teacher and student, utilizing its own outputs rather than relying on posterior distributions. As a result, it is often considered a discriminative self-supervised objective \cite{andonian2022robust}, \cite{dong2023maskclip}, \cite{ji2021refine}, \cite{kim2021self}. Under this mechanism, the network learns from its historical iterations or differently perturbed views, progressively refining its high-dimensional representations.
By eliminating the prerequisite to pre-train and freeze a massive external teacher network, self-distillation significantly reduces the memory and time overhead. More importantly, it typically enhances the model's robustness \cite{Data2vec}, \cite{caron2021emerging}, \cite{cheng2021data}, \cite{li2021align}.
Motivated by the self-distillation, in the proposed MMAE framework, we seamlessly integrate the self-distillation mechanism between a twin of MAEs to achieve interactive pre-training. 

\section{The Proposed Mean Masked AutoEncoder}
An overview of our MMAE is illustrated in Fig.~\ref{MMAE_overview}, which contains a traffic preprocessing unit (d), a FlowMix unit (a), a student MAE (b), a teacher MAE (c), and a Statistics-based Flow Matcher (e) for constructing flow pairs. 

\subsection{Traffic Preprocessing}
\begin{figure*}[t]
\begin{center}
\includegraphics[width=1.0\linewidth]{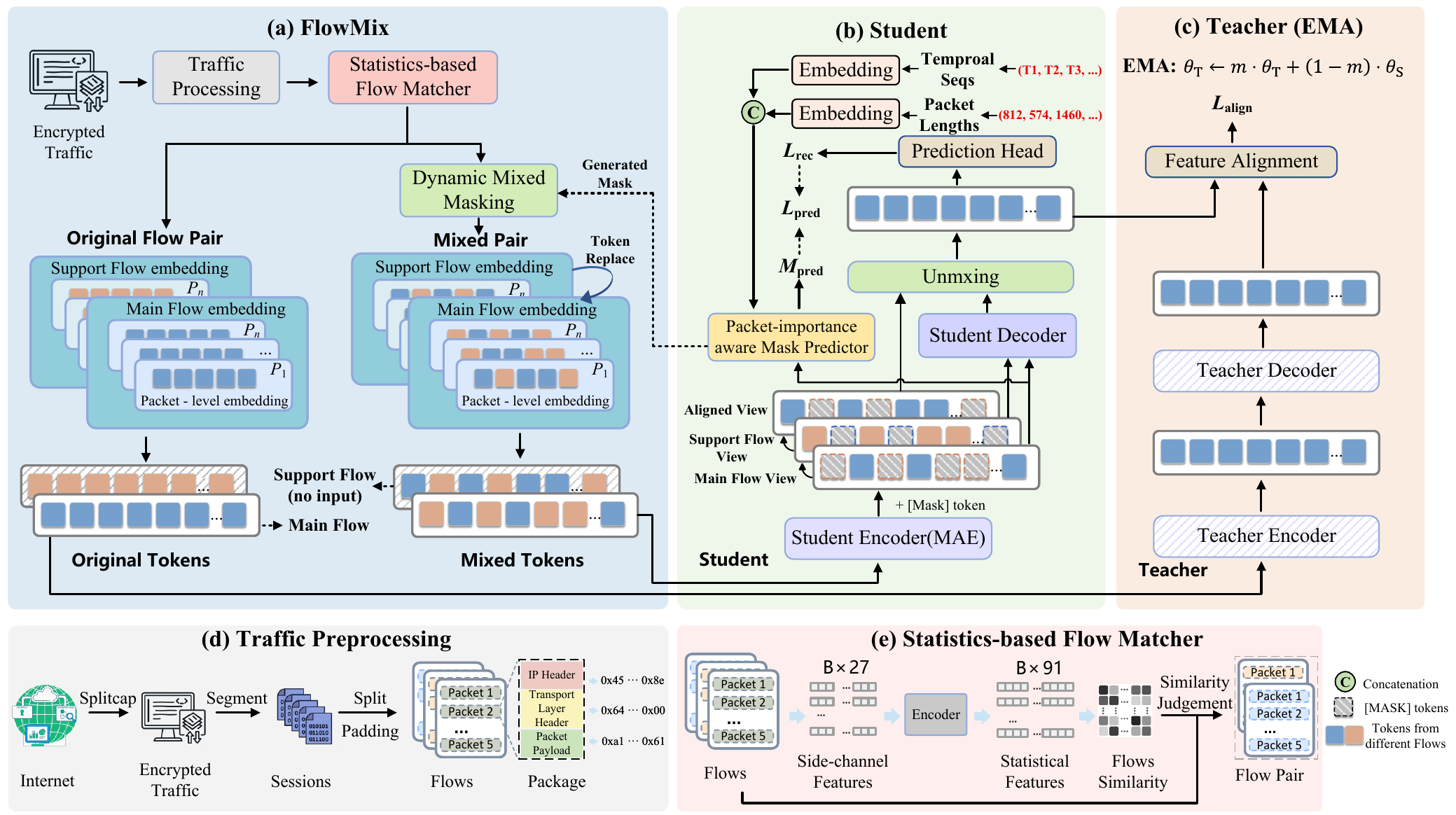}
\end{center}
   \caption{Flowchar of the proposed MMAE, which mainly includes FlowMix and teacher-student based twin MAEs. 
   MMAE incorporates a traffic pre-processing unit, a statistics-based flow matcher and a cross-flow mixing unit (i.e., FlowMix). Within the self-distillation teacher-student twin MAE architecture, flow-level reference features are exploited as supervision for reconstruction and mask prediction tasks.}
\label{MMAE_overview}
\end{figure*}

To transform the heterogeneous and variable-length raw network packets into representations suitable for models, Fig. \ref{MMAE_overview} (d) illustrates the complete preprocessing workflow.
First, we use the SplitCap tool to segment the PCAP files captured from network interfaces into independent session flows based on the five-tuple rule. To eliminate potential model bias towards specific hosts, the IP addresses and port details information within the flows are anonymized.
Subsequently, we apply uniform truncation and padding to the packets, fixing each packet to 320 bytes (80 bytes of protocol header features and 240 bytes of payload). This specific allocation preserves critical initial information from both the control and application layers while avoiding interference from excessive zero-padding.

We extract the first 5 packets from each standardized session. If a session contains fewer than 5 packets, appropriate zero-padding is applied. These extracted packets are then concatenated to form a one-dimensional integer array with a fixed length of $L=1600$ bytes. Finally, to improve the model's convergence and numerical stability, each byte value is normalized from the original $[0, 255]$  to $[0, 1]$ interval.

\begin{table}[t]
    \centering
    \caption{Summary of Extracted Side-Channel Statistical Features}
    \label{side_channel_features}
    \renewcommand{\arraystretch}{1}
    \footnotesize
    \begin{tabularx}{\columnwidth}{@{} p{1.2cm} p{2.1cm} X @{}}
        \toprule
        \textbf{Category} & \textbf{Feature Name} & \textbf{Description} \\
        \midrule
        
        \multirow{6}{*}{\shortstack[l]{\textbf{Time}\\(6 dims)}} 
        & F-Duration & Total time of a flow from start to end \\
        & F-Time & Timestamp of the first packet in a flow \\
        & P-inter-min & Shortest time interval between packets \\
        & P-inter-max & longest time interval between packets \\
        & P-inter-avg & Average deviation of packet intervals \\
        & P-inter-std & Standard deviation of packet intervals \\
        \midrule
        
        \multirow{11}{*}{\shortstack[l]{\textbf{Size}\\(11 dims)}} 
        & P-total & Total number of packets in a flow \\
        & Downlink-Bytes & Total bytes transmitted downlink \\
        & Downlink-Count & Total packets count transmitted downlink \\
        & Payload-P-Count & Number of packets with payload data \\
        & Payload-P-min & Smallest payload sizes \\
        & Payload-P-max & Largest payload sizes \\
        & Payload-P-std & Standard deviation of payload sizes \\
        & P-Length-min & Smallest and largest packet lengths \\
        & P-Length-max & Smallest and largest packet lengths \\
        & P-Length-avg & Average deviation of packet lengths \\
        & P-Length-std & Standard deviation of packet lengths \\
        \midrule
        
        \multirow{10}{*}{\shortstack[l]{\textbf{Flag}\\(10 dims)}} 
        & TCP Count & Number of TCP packets in a flow \\
        & UDP Count & Number of UDP packets in a flow \\
        & DNS Count & Number of DNS packets in a flow \\
        & ICMP Count & Number of ICMP packets in a flow \\
        & SYN Count & Number of SYN flags in a flow \\
        & FIN Count & Number of FIN flags in a flow \\
        & ACK Count & Number of ACK flags in a flow \\
        & PSH Count & Number of PSH flags in a flow \\
        & URG Count & Number of URG flags in a flow \\
        & RST Count & Number of RST flags in a flow \\
        
        \bottomrule
    \end{tabularx}
\end{table}

\subsection{FlowMix Mechanism} \label{section:CDM}
To overcome the limited contextual information within a single flow, we introduce a dynamic Flow Mixing strategy to enhance the model's flow-level discriminative capabilities with perturbation tokens. However, random pairwise mixing of traffic flows have drastically different behavioral patterns. In such cases, the resulting reconstruction task becomes simple, as the model can easily distinguish them, failing to learn deep semantics. Therefore, effective cross-flow mixing requires identifying traffic pairs with similar patterns.

Side-channel statistical features are inherent physical attributes of network traffic, which remain unaffected by encryption and can therefore intuitively reflect flow-level patterns, packet-level dynamics, and the overall communication environment. Statistics-based Flow Matcher (SFM) leverages these statistical priors as anchors to reliably estimate the physical similarity between flows, which establishes a principled foundation for the mixing process.

\subsubsection{Statistics-based Flow Matcher}
As illustrated in Fig. \ref{MMAE_overview} (e), SFM extracts 27 highly expressive side-channel statistical features from flow-level PCAP files. These features encompass packet-level behaviors, temporal patterns, and protocol interaction modes. Based on their primary physical implications, we categorize these attributes into three principal groups: \textit{Time} Features, \textit{Size} Features, and \textit{Flag \& Protocol} Features.
\begin{itemize}
    \item
    \textbf{Time Features:} These features describe the temporal patterns of packet arrivals and session persistence. For example, real-time communication applications typically exhibit stable and frequent inter-arrival times, while non-real-time applications show bursty time intervals.
    
    \item
    \textbf{Size Features:} These attributes capture the distributional patterns of data volume transmission across the network flow, including the statistical regularity of overall packet length, payload size, and downlink traffic distribution.
    
    \item
    \textbf{Flag \& Protocol Features:} These features record the trigger frequency of control flags (e.g., SYN, FIN, PSH), which provide intuitive insights into the connection management strategies. For example, applications that frequently establish connections generate a large number of SYN/FIN flags, while persistent connection applications rarely generate such flags.
\end{itemize}

The specific names and physical meanings of the 27 side-channel statistical features are shown in Table \ref{side_channel_features}.

Subsequently, an encoder maps the raw features to a higher-order abstract representation. We denote the input raw statistical feature matrix as $X_{stat} \in \mathbb{R}^{1 \times 27}$. The entire feature encoding process can be represented as follows:
\begin{equation}
\begin{aligned}
\label{eq:SFM_encoder}
X_{norm}&=\mathrm{BatchNorm}(X_{stat})\\
F_{cnn}&=\mathrm{MaxPool}\left(\mathrm{CNN1D}(X_{norm})\right)\\
F_{deep}&=\mathrm{MLP}(F_{cnn})\\
F_{out}&=\mathrm{Concat}(X_{norm},F_{deep})
\end{aligned}
\end{equation}
where $X_{norm} \in \mathbb{R}^{1 \times 27}$ denotes the features processed via 1D Batch Normalization. $\mathrm{CNN1D}(\cdot)$ represents a deep network comprising two 1D convolutional layers. Subsequently, an adaptive max-pooling layer $\mathrm{MaxPool}(\cdot)$ is used for feature compression to extract the most salient sequence features $F_{cnn}$. Finally, the pooled features are mapped to a high-order deep representation $F_{deep} \in \mathbb{R}^{1 \times 64}$.
To avoid losing the explicit semantics of the original physical properties, the encoder employs a residual concatenation operation, resulting in a $91$-dimensional fused feature $F_{out}$.

Next, we must precisely pair each main flow with a corresponding support flow to construct the mixing input. To this end, we construct symmetric pairs using a greedy matching \cite{preis1999linear} strategy within a mini-batch, which is detailed in Algorithm \ref {symmetric_pairing}. The final output of this algorithm is an index vector $P$, which encapsulates the symmetric pairing assignments for the entire batch. 

\subsubsection{Dynamic Mixed Masking} \label{section:DMM}

After selecting the paired flows through SFM, executing an effective mixing strategy is crucial. To avoid the triviality of blindly swapping random tokens with varying information densities, we design the Dynamic Mixed Masking (DMM) module. Instead of the conventional random masking in standard MAE, we guide the selection process using the prediction loss $L_{pred}$ from the student model, which is elaborated as the Packet-importance aware Mask Predictor (PMP) in Section~\ref{section:PMP}. Regions with higher prediction errors indicate dense semantics that the model struggles to learn. Therefore, we prioritize masking and replacing these challenging regions with interfering features from paired flows indexed by $P$ from SFM module.

Specifically, we partition the input traffic sequence into $N_x = 400$ patches with a step size of 4, and linearly project them into $D = 256$ dimensions. This yields the main flow's token sequence, denoted as $X_{main} \in \mathbb{R}^{{N_x} \times D}$. Given masking ratio $r_{mask}$, the number of tokens to be replaced by support flow features $X_{supp}$ corresponding to paired flows chosen by SFM is $N_{mask} = \lfloor {N_x} \cdot r_{mask} \rfloor$, leaving $N_{keep} = {N_x} - N_{mask}$ tokens retained, where $\lfloor \cdot \rfloor$ represents the operation of rounding down. To progressively increase the difficulty of the pretext task, the total masked tokens $N_{mask}$ are divided into replaced parts with the hard mask ratio of $r_{hard}$ and random mask parts with a ratio of $r_{rand}$, respectively.

Based on the predicted loss $\mathcal{L}_{pred} \in \mathbb{R}^{N_x}$ from PMP, we identify the indices $\Omega_{hard}$ of the $K_{hard}$ most challenging patches via the $TopK\_Indices$ operation:
\begin{equation}
\begin{aligned}
\label{eq:DMM_index}
\Omega_{hard} = \text{TopK\_Indices}(\mathcal{L}_{pred}, K_{hard})
\end{aligned}
\end{equation}

To efficiently execute both random and targeted masking simultaneously, we first generate uniform noise $\mathcal{N} \sim U(0, 1)^{N_x}$, and reformulate each value $\mathcal{N}_{i,j}$ as: 
\begin{equation}
\begin{aligned}
\hat{\mathcal{N}} = \begin{cases} \mathcal{N}_{i,j} + \eta, & \text{if } j \in \Omega_{hard} \\ \mathcal{N}_{i,j}, & \text{otherwise} \end{cases}
\end{aligned}
\end{equation}
given that $\eta \gg 1$, the values of selected challenging patches in $\hat{\mathcal{N}}$ become significantly larger than the uniform noise. By sorting $\hat{\mathcal{N}}$ in ascending order along the sequence dimension, we construct the binary mask matrix $M \in \{0, 1\}^{N_x}$:
\begin{equation}
\begin{aligned}
M_j = \begin{cases} 0, & \text{if } j < N_{keep} \quad (\text{Keep}) \\ 1, & \text{if } j \ge N_{keep} \quad (\text{Replace}) \end{cases}
\end{aligned}
\end{equation}
where $j$ is the index in the sorted sequence, and $N_{keep}$ is the target number of retained tokens.

However, during the early stages of training, the model has not yet extracted meaningful feature representations, and $L_{pred}$ is highly uncertain. Therefore, we formulate the ratio of $\mathcal{L}_{pred}$-guided hard patches as a dynamic scaling function that evolves with the training epoch. This design fundamentally incorporates the paradigm of Curriculum Learning.
For the current training epoch $t$, the definition is as follows:
\begin{equation}
\begin{aligned}
\label{eq:DMM_ratio}
r_{hard}(t) = R_{max} \cdot \left( \frac{t}{T_{total}} \right)
\end{aligned}
\end{equation}
as $t$ increases from $0$ to $T_{total}$, the ratio $r_{hard}(t)$ scales linearly up to its maximum threshold $R_{max}$.

\begin{algorithm}[t]
\caption{Symmetric Pairing Construction}
\label{symmetric_pairing}
\renewcommand{\algorithmiccomment}[1]{\quad\(\triangleright\) #1}
\begin{algorithmic}[1] 
\Require Batch side-channel statistical feature matrix $F_{out} \in \mathbb{R}^{B \times D}$ (Batch size $B$, Feature dimension $D=91$)
\Ensure Symmetric pairing index vector $P \in \mathbb{Z}^{B}$

\State $F_{norm} \leftarrow \frac{F_{out}}{\|F_{out}\|_2}$ \Comment{Row-wise L2 normalization}
\State $\mathcal{S} \leftarrow F_{norm} \cdot F_{norm}^\top$ 
\For{$i = 1$ \textbf{to} $B$}
    \State $\mathcal{S}[i, i] \leftarrow -\infty$ \Comment{Prevent self-pairing}
\EndFor
\State $P \leftarrow [1, 2, \dots, B]$ \Comment{Initialize indices}
\State $\mathcal{S}_{current} \leftarrow \mathcal{S}$

\For{$step = 1$ \textbf{to} $\lfloor B / 2 \rfloor$}
    \State $row, col \leftarrow argmax_{i, j}(\mathcal{S}_{current})$ \Comment{Maximum coordinates}
    \If{$\mathcal{S}_{current}[row, col] = -\infty$}
        \State \textbf{break} \Comment{Terminate if all elements are masked}
    \EndIf
    
    \State $P[row] \leftarrow col$
    \State $P[col] \leftarrow row$  \Comment{Symmetric bi-directional mapping}
    
    \Statex \hspace{1.5em} \Comment{Exclude paired rows and columns}
    \State $\mathcal{S}_{current}[row, :] \leftarrow -\infty$; \quad $\mathcal{S}_{current}[:, row] \leftarrow -\infty$
    \State $\mathcal{S}_{current}[col, :] \leftarrow -\infty$; \quad $\mathcal{S}_{current}[:, col] \leftarrow -\infty$
\EndFor

\State \Return $P$
\end{algorithmic}
\end{algorithm}

Finally, utilizing the pairing index vector $P$ generated by Algorithm~\ref{symmetric_pairing}, we extract the corresponding support flow token sequences $X_{supp}$ within the batch. The final flow mixing sequence $X_{mix}$ is:

\begin{equation}
\begin{aligned}
\label{eq:DMM_mixing}
X_{mix} = X_{main} \odot (1 - M) + X_{supp} \odot M
\end{aligned}
\end{equation}
where $\odot$ denotes the Hadamard product. When a value in $M$ is 0, the token retains the main flow feature $X_{main}$. Otherwise, it is replaced by a feature from the support flow $X_{supp}$.
Notably, the support flow serves only as a repository of distorted features. It is not fed into the network as an independent sequence as indicated by the ``no input" state in Fig. \ref{MMAE_overview} (a). Thus, the core entity of the pre-training task is always based on the main flow.

To preserve the positional relationships of packets and bytes in traffic, we add positional encoding $E_{pos} \in \mathbb{R}^{N \times D}$ to the sequence, where $N = N_x + 1$ is the total sequence length. Following the standard Transformer architecture, we also prepend a learnable class token $x_{cls} \in \mathbb{R}^{1 \times D}$. Finally, the mixed feature embedding fed into the student model is formulated as:
\begin{equation}
\begin{aligned}
\label{eq:student_input}
X^{student} = [x_{cls}; X_{mix}] + E_{pos}
\end{aligned}
\end{equation}
where $X_{mix} \in \mathbb{R}^{N_x \times D}$ represents the cross-flow mixed feature and $[;]$ represents the concatenation operation.

The unmasked raw feature fed into teacher model is:
\begin{equation}
\begin{aligned}
\label{eq:DMM_teacher_input}
X^{teacher} = [x_{cls}; X_{main}] + E_{pos}
\end{aligned}
\end{equation}

\subsection{Student Branch}
The core objective of the student branch is to extract representative features from the cross-flow mixed sequence $X^{student}$ by reconstructing the main flow feature. 

First, the student encoder $\mathcal{E}_S$ processes the mixed token sequence $X^{student}$. Through multi-layer self-attention, the features of the main and support flows are deeply integrated:
\begin{equation}
\begin{aligned}
\label{eq:PMP_latent}
Z^{student} = \mathcal{E}_S(X^{student})
\end{aligned}
\end{equation}
were $Z^{student} \in \mathbb{R}^{N \times D}$ represents the hidden-layer feature. This representation is then fed into its decoder and another parallel branch, i.e., PMP. In the following, we first detail the decoder and its reconstruction mechanism.

\subsubsection{Dual-view Construction and Decoding}
To separately recover the semantics of the main and support flows from the mixed tokens $Z^{student}$, we construct dual views before the decoder. Using the binary mask $M \in \{0, 1\}^{N_x}$ generated in the DMM, we obtain the main flow view $V_{main}$ and its symmetric support flow view $V_{supp}$. Specifically, $V_{main}$ retains the main flow tokens and replaces the positions of the support flow with learnable mask token $E_{[MASK]}$:
\begin{equation}
\begin{aligned}
\label{eq:PMP_main_view}
V_{main} = Z^{student} \odot (1 - M) + E_{[MASK]} \odot M
\end{aligned}
\end{equation}
Conversely, $V_{supp}$ is represented as:
\begin{equation}
\begin{aligned}
\label{eq:PMP:supp_view}
V_{supp} = Z^{student} \odot M + E_{[MASK]} \odot (1 - M)
\end{aligned}
\end{equation}

These two views are concatenated along the batch dimension and fed into the decoder $\mathcal{D}_S$ for reconstruction:
\begin{equation}
\begin{aligned}
\label{eq:decoder}
H_{main} &= \mathcal{D}_S(V_{main} + E_{pos}^{dec}) \\
H_{supp} &= \mathcal{D}_S(V_{supp} + E_{pos}^{dec})
\end{aligned}
\end{equation}
where $H_{main} \in \mathbb{R}^{N_x \times D}$ and $H_{supp} \in \mathbb{R}^{N_x \times D}$ represent the reconstructed main and support view tokens, respectively. For brevity, the class token in the output sequence is omitted here.

\begin{figure*}[t]
\begin{center}
\includegraphics[width=1.00\linewidth]{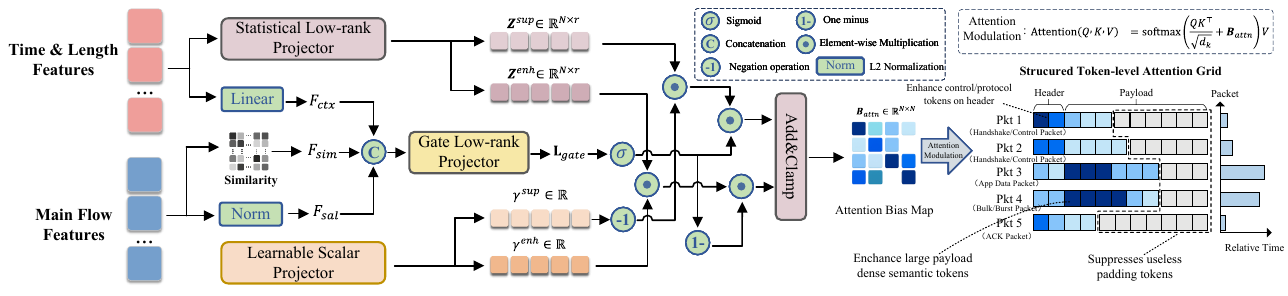}
\end{center}
   \caption{Architecture of the Packet-importance aware Mask Predictor (PMP). PMP leverages packet-level side-channel priors to generate a low-rank attention bias via a token-aware gating mechanism. This bias dynamically modulates self-attention to identify and mask challenging, semantically dense regions.}
\label{loss_pres}
\vspace{-0.3cm}
\end{figure*}

\subsubsection{Unmixing and Reconstruction}
$H_{main}$ and $H_{supp}$ contain the predictions for the masked regions. To recover the complete semantics of the main flow, we perform unmixing and reconstruction. First, we remap the index of support view token $H_{supp}$ back to their original batch positions according to the index vector $P$ from SFM in Algorithm~\ref{symmetric_pairing}, obtaining the aligned token:
\begin{equation}
\begin{aligned}
\label{eq:unmixing_begin}
H_{aligned} = H_{supp}[P]
\end{aligned}
\end{equation}


The unmixing process is then formulated as:
\begin{equation}
\begin{aligned}
\label{eq:H_unmix}
H_{unmix} = H_{main} \odot M + H_{aligned} \odot (1 - M)
\end{aligned}
\end{equation}

Notably, $H_{unmix}$ restores the original structure of the main flow, utilized for global semantic alignment with the reconstructed full tokens by the teacher network.
Following unmixing, the model performs fine-grained byte-level reconstruction and computes the reconstruction loss:
\begin{equation}
\begin{aligned}
\label{eq:l_res}
\mathcal{L}_{rec} = MSE(H_{unmix}, {X}_{main})
\end{aligned}
\end{equation}
where 
$MSE(\cdot)$ indicates Mean Squared Error. Since the model predicts masks from dual views simultaneously in a single forward pass, the MSE can be directly calculated across all tokens to fully optimize the network.

\subsection{Packet-importance aware Mask Predictor}
\label{section:PMP}
To effectively identify the most informative patches, we introduce a Packet-importance aware Mask Predictor (PMP). Packet-level side-channel features are natural indicators of reconstruction difficulty. Inter-arrival time ($S_{time}$) reflects macroscopic burstiness, while payload length ($S_{len}$) determines microscopic information capacity. Leveraging these multi-granularity priors, PMP dynamically allocates attention to accurately predict reconstruction difficulty. As shown in Fig. \ref{loss_pres}, this is achieved by generating an attention bias.

We first broadcast the packet-level statistical signal $S_{time}$ and $S_{len}$ along the byte dimension to match the original sequence length. Then, respective 1D convolutions are applied to map them into the token latent space. The resulting embeddings $E_t$ and $E_l$, are concatenated and fused to form the statistical context feature $Z_{stat} \in \mathbb{R}^{N_x \times D_{stat}}$:
\begin{equation}
\begin{aligned}
\label{eq:attn_bias_begin}
E_t &= \text{Conv1D}(\text{Broadcast}(S_{time})) + \text{PE}_t \\
E_l &= \text{Conv1D}(\text{Broadcast}(S_{len})) + \text{PE}_l \\
Z_{stat} &= \text{LayerNorm}(W_{fuse}[E_t; E_l])
\end{aligned}
\end{equation}
where $\text{PE}_t$ and $\text{PE}_l$ denote the respective positional encodings, and $W_{fuse}$ is a learnable weight matrix used for feature fusion.

To modulate the attention distribution, PMP generates suppressive and enhancing matrices, defined by a policy indicator $k \in \{sup, enh\}$. For each policy, the statistical feature $Z_{stat}$ is projected into a rank-$r$ space via learnable matrices $W_U^k$ and $W_Y^k \in \mathbb{R}^{D_{stat} \times r}$, producing low-rank factors.
\begin{equation}
\begin{aligned}
\mathbf{U}^k = Z_{stat} W_U^k, \quad \mathbf{Y}^k = Z_{stat} W_Y^k
\end{aligned}
\end{equation}
The dense bias matrix $Z^k \in \mathbb{R}^{N_x \times N_x}$ is then computed via inner product:
\begin{equation}
\begin{aligned}
Z^k = \frac{1}{\sqrt{r}} (\mathbf{U}^k)(\mathbf{Y}^k)^\top
\end{aligned}
\end{equation}

Since different tokens play varying roles in traffic contexts, a gating mechanism dynamically decides whether to suppress or enhance each position. We extract three metrics for the $i$-th token: a similarity feature $F^i_{sim}$, a saliency feature $F^i_{sal}$, and a statistical context feature $F^i_{ctx}$:
\begin{equation}
\begin{aligned}
  F^i_{sim} &= \frac{1}{N_x} \sum_j \text{ReLU}(\cos(v_i, v_j)) \\
  F^i_{sal} &= \text{Linear}(\|v_i\|_2) \\
  F^i_{ctx} &= \text{Linear}(Z_{stat,i})
\end{aligned}
\end{equation}
where $v_i \in \mathbb{R}^D$ denotes the $i$-th token in the main flow view $V_{main}$. These metrics are concatenated to form the comprehensive gated features $F^i_{gate} =[F^i_{sim}; F^i_{sal}; F^i_{ctx}]$. Subsequently, gating logits $\mathbf{L}_{gate}$ are generated using a similar matrix low-rank projection parameterized by learnable weight matrices $\mathbf{W}_a$ and $\mathbf{W}_b$. These logits are then used to calculate the suppression probability $p^{sup}$ and enhancement probability $p^{enh}$ for each position:
\begin{equation}
\begin{aligned}
\mathbf{L}_{gate} &= (F_{gate} \mathbf{W}_a)(F_{gate} \mathbf{W}_b)^\top \\
p^{sup} &= \text{Sigmoid}(\mathbf{L}_{gate}) \\
p^{enh} &= 1 - p^{sup}
\end{aligned}
\end{equation}

Finally, the suppressive and enhancing bias matrices are weighted by the calculated gating probabilities. To ensure numerical stability, the final attention bias matrix $\boldsymbol{B}_{attn}$ is clamped to the interval $[-5, 5]$:
\begin{equation}
\begin{split}
\boldsymbol{B}_{attn} = \text{Clamp}\Big(
p^{sup} \odot (- \text{Softplus}(\gamma^{sup}) \cdot Z^{sup}) \\
+ p^{enh} \odot (\text{Softplus}(\gamma^{enh}) \cdot Z^{enh}), -5, 5 \Big)
\end{split}
\end{equation}
where $\gamma^{sup}$ and $\gamma^{enh}$ are learnable scalars controlling the bias magnitudes. 
In the implementation, PMP processes $Z^{student}$ through a dedicated prediction network $\mathcal{F}_{PMP}$ to output the patch-wise reconstruction difficulty prediction:
\begin{equation}
\begin{aligned}
M_{pred} &= \mathcal{F}_{PMP}(Z^{student})
\end{aligned}
\end{equation}
where $M_{pred} \in \mathbb{R}^{N_x \times D}$, and $\mathcal{F}_{PMP}$ is implemented in the structure of a 2-layer Transformer block. Within $\mathcal{F}_{PMP}$, the generated bias matrix $\boldsymbol{B}_{attn}$ is injected directly into the attention score computation:
\begin{equation}
\begin{aligned}
\label{eq:attn_with_bias}
\text{Attention}(Q, K, V) &= \text{softmax} \left( \frac{Q K^\top}{\sqrt{d_k}} + \boldsymbol{B}_{attn} \right) V
\end{aligned}
\end{equation}

The prediction $M_{pred}$ is subsequently fed back to the module described in Section~\ref{section:CDM} to guide the dynamic mask generation of cross-flow samples for the next training iteration.

PMP is supervised by the reconstruction loss $\mathcal{L}_{rec}$. However, as pre-training progresses, the absolute reconstruction loss undergoes drastic decay and becomes noisy. To circumvent this instability, we adopt a pairwise ranking loss inspired by RankNet \cite{ranknet}. For any two patches $i$ and $j$, we construct binary ranking labels based on $\mathcal{L}_{rec}$:
\begin{equation}
\begin{aligned}
\label{eq:l_pred_ranking}
\mathbf{I}_{i,j}^+ &= \begin{cases} 1, & \text{if } \mathcal{L}_{rec}^{(i)} > \mathcal{L}_{rec}^{(j)} \\ 0, & \text{otherwise} \end{cases} \\
\mathbf{I}_{i,j}^- &= \begin{cases} 1, & \text{if } \mathcal{L}_{rec}^{(i)} < \mathcal{L}_{rec}^{(j)} \\ 0, & \text{otherwise} \end{cases}
\end{aligned}
\end{equation} We then compute the prediction difference $\Delta_{i,j} = M_{pred}^{(i)} - M_{pred}^{(j)}$ and optimize this relative ranking using binary cross-entropy (BCE):
\begin{equation}
\label{eq:l_pred}
\mathcal{L}_{pred}
=
\mathrm{BCE}\big(
\sigma(\Delta_{i,j}),
\mathbf{I}_{i,j}^{+}
\big)
+
\mathrm{BCE}\big(
1-\sigma(\Delta_{i,j}),
\mathbf{I}_{i,j}^{-}
\big)
\end{equation}
where $\sigma(\cdot)$ is the Sigmoid activation function.

\begin{algorithm}[t]
\small
\caption{Training Pipeline of MMAE}
\label{alg:mta_mae}
\begin{algorithmic}[1]
\Require Unlabeled traffic data $\mathcal{D}_{unlabel}$, labeled traffic data $\mathcal{D}_{label}$, total pre-training epochs $T_{total}$, fine-tuning epochs $E_{fine}$, initial student parameters $\theta_S$, EMA momentum $m$.

\State Initialize Teacher parameters $\theta_T$ using $\theta_S$

\State \textbf{Phase 1: Pre-training}
\For{$t = 1$ \textbf{to} $T_{total}$}
    \For{each mini-batch in $\mathcal{D}_{unlabel}$}
        \State \textcolor{green!70!black}{\textit{\# Flow Mixing}}
        \State Extract and encode side-channel statistical features (Eq.~\ref{eq:SFM_encoder})
        \State Construct symmetric pairing index $P$ (Algorithm~\ref{symmetric_pairing})
        \State Generate dynamic mixed mask $M$ (Eq.~\ref{eq:DMM_index}-\ref{eq:DMM_ratio})
        \State Construct cross-flow mixed input $X_{student}$ (Eq.~\ref{eq:DMM_mixing}, \ref{eq:student_input})
        \State Construct unmasked input $X_{teacher}$ (Eq.~\ref{eq:DMM_teacher_input})
        
        \State \textcolor{green!70!black}{\textit{\# Student Branch}}
        \State Construct and decode dual views $V_{main}, V_{supp}$ (Eq.~\ref{eq:PMP_latent}-\ref{eq:decoder})
        \State Unmix features to obtain $H_{unmix}$ (Eq.~\ref{eq:unmixing_begin}, \ref{eq:H_unmix})
        \State Compute reconstruction loss $\mathcal{L}_{rec}$ (Eq.~\ref{eq:l_res})
        \State \textcolor{green!70!black}{\textit{\# Packet-importance aware Mask Predictor}}
        \State Generate attention bias $\boldsymbol{B}_{attn}$ (Eq.~\ref{eq:attn_bias_begin}-\ref{eq:attn_with_bias})
        \State Compute mask prediction loss $\mathcal{L}_{pred}$ (Eq.\ref{eq:l_pred_ranking}, ~\ref{eq:l_pred})
        
        \State \textcolor{green!70!black}{\textit{\# Teacher Branch with Self-Distillation}}
        \State Extract global contextual features $H_{teacher}$ (Eq.~\ref{eq:teacher_feature})
        \State Compute semantic alignment loss $\mathcal{L}_{align}$ (Eq.~\ref{eq:l_align})
        
        \State \textcolor{green!70!black}{\textit{\# Model Optimization}}
        \State Compute total pre-training loss $\mathcal{L}_{pre}$ (Eq.~\ref{eq:pretrain_loss})
        \State Update Student parameters $\theta_S \leftarrow \theta_S - \eta \nabla_{\theta_S} \mathcal{L}_{pre}$
        \State Update Teacher parameters $\theta_T$ via EMA (Eq.~\ref{eq:ema})
    \EndFor
\EndFor

\State \textbf{Phase 2: Fine-tuning}
\State \textcolor{green!70!black}{\textit{\# Downstream Initialization}}
\State Retain pre-trained student encoder $\theta_S$ and initialize downstream classification head
\For{epoch $= 1$ \textbf{to} $E_{fine}$}
    \For{each batch $(X, y)$ in $\mathcal{D}_{label}$}
    \State \textcolor{green!70!black}{\textit{\# Supervised Optimization}}
        \State Extract features using student encoder and predict probability distribution $\hat{y}$
        \State Compute fine-tuning loss $\mathcal{L}_{CE}$ (Eq.~\ref{eq:fine_tuning_loss})
        \State Update student encoder and classification head parameters by minimizing $\mathcal{L}_{CE}$
    \EndFor
\EndFor

\State \Return Fine-tuned model parameters
\end{algorithmic}
\end{algorithm}

\subsection{Teacher Branch with Self-Distillation}
During the reconstruction phase, the student model inputs only severely disrupted mixed fragments. In contrast, the teacher model processes the complete global flow, extracting high-level semantic representations as supervision signals.

The teacher branch takes the original feature embedding $X^{teacher}$ constructed in Section \ref{section:CDM} as input to extract the full contextual feature $H_{teacher}$:
\begin{equation}
\begin{aligned}
\label{eq:teacher_feature}
H_{teacher} = \mathcal{D}_T \left( \mathcal{E}_T (X^{teacher}) \right)
\end{aligned}
\end{equation} where $\mathcal{E}_T$ and $\mathcal{D}_T$ denote the encoder and decoder of the teacher, respectively.

To guide the student in learning global traffic semantics within the latent space, we introduce a feature alignment mechanism. Specifically, we align the unmixed feature $H_{unmix}$ from the student with the $H_{teacher}$ from the teacher using the alignment loss $\mathcal{L}_{align}$:
\begin{equation}
\begin{aligned}
\label{eq:l_align}
\mathcal{L}_{align} = 1 - \text{sim}(H_{unmix}, H_{teacher})
\end{aligned}
\end{equation}
where $\text{sim}(\cdot)$ denotes the cosine similarity.
By minimizing $\mathcal{L}_{align}$, the student network is subjected to a strict constraint: \textit{even when the student's input is heavily contaminated by other flow, its unmixed representation must align with the teacher's view of the uncorrupted context}. This semantic distillation improves model's robustness against traffic obfuscation.

Notably, the teacher's weights are updated using an Exponential Moving Average (EMA) strategy:
\begin{equation}
\begin{aligned}
\label{eq:ema}
\theta_T \leftarrow m \cdot \theta_T + (1 - m) \cdot \theta_S
\end{aligned}
\end{equation}
where $\theta_T$ and $\theta_S$ denote the parameters of the teacher and student models, respectively. $m \in[0, 1]$ is the momentum coefficient controlling the update rate.
In this paper, the momentum $m$ is scheduled via cosine annealing:
\begin{equation}
m_t = m_{final} + \frac{1}{2} (m_{base} - m_{final}) \left( 1 + \cos\left( \frac{\pi t}{T} \right) \right)
\end{equation}
where $t$ and $T$ denote the current and total training steps, respectively. $m$ is smoothly increased from the base value $m_{base}$ to the final value $m_{final}$, ensuring model stability. 

Since the teacher and student are a twin of MAE, and the teacher is updated through EMA strategy, we therefore call the model \textbf{Mean MAE}.

\subsection{Model Optimization}

\textbf{Pre-training phase.} The overall pre-training objective is formulated as:
\begin{equation}
\begin{aligned}
\label{eq:pretrain_loss}
\mathcal{L}_{pre} = \mathcal{L}_{rec} + \lambda_1 \mathcal{L}_{pred} + \lambda_2 \mathcal{L}_{align}
\end{aligned}
\end{equation}
where $\lambda_1$ and $\lambda_2$ are balancing hyperparameters.

\textbf{Fine-tuning phase.} For downstream tasks, we fine-tune the full model by retaining only the pre-trained student encoder as the backbone. The extracted features are fed into a classification head to predict the probability distribution $\hat{y} \in \mathbb{R}^c$, where $c$ is the number of traffic categories. The model is optimized using the Cross-Entropy loss between the ground-truth labels $y$ and predictions $\hat{y}$:
\begin{equation}
\begin{aligned}
\label{eq:fine_tuning_loss}
\mathcal{L}_{CE} = \text{CrossEntropy}(y, \hat{y})
\end{aligned}
\end{equation}

To provide a comprehensive overview, the training pipeline of the proposed MMAE is detailed in Algorithm \ref{alg:mta_mae}.

\section{Experiments}
\subsection{Experimental Setup}
\subsubsection{Datasets and Preprocessing}

For the encrypted traffic classification task, the model is pre-trained using six public real-world datasets: ISCXVPN 2016 \cite{ISCX2016}, ISCXTor 2016 \cite{ISCX2016}, CrossPlatform (Android) \cite{Flowprint}, CrossPlatform (iOS) \cite{Flowprint}, USTC-TFC \cite{ustc}, and CICIoT2022 \cite{ciciot}. During the fine-tuning phase, we evaluate the model on these six datasets and the CSTNET-TLS 1.3 dataset \cite{Et_bert}. Each dataset is split into training, validation, and test sets at an 8:1:1 ratio.
\begin{itemize}
    \item
    \textbf{ISCXVPN2016:} This dataset comprises communication application traffic captured by the Canadian Institute for Cybersecurity, which contains VPN and non-VPN traffic from 16 applications categorized into 7 types, yielding 16,048 processed samples.
    \item
    \textbf{ISCXTor2016:} This dataset contains application traffic using Tor for encrypted communication, which includes traffic data from 8 communication categories and totally 14,569 samples.
    \item
    \textbf{USTC-TFC2016:} This dataset consists of encrypted traffic from 20 application types, evenly split between 10 benign and 10 malicious categories, comprising 50,677 processed samples.
    \item
    \textbf{CrossPlatform (Android \& iOS):} These datasets include encrypted traffic from the top 100 Android and iOS apps in the US, China, and India. To prevent long-tail classes from degrading model performance, we discarded categories with fewer than 50 samples, resulting in 181 and 124 categories, which comprise 54,011 and 48,787 processed samples, respectively.
    \item
    \textbf{CICIoT2022:} This dataset is collected from a laboratory network, designed for IoT vulnerability testing, encompassing six mainstream IoT malicious attack categories, yielding 22,634 samples. 
    \item
    \textbf{CSTNET-TLS 1.3:} This dataset contains encrypted traffic data over the encryption protocol TLS 1.3, which is one of the most cutting-edge, widely used, and thoroughly encrypted transport protocol standards on the internet. This data was captured from the China Science and Technology Network backbone across 120 applications, closely reflecting actual ISP traffic distributions, contributing 46,356 processed samples.
\end{itemize}

\begin{table*}[t]
\centering
\caption{Comparisons with previous SoTA methods on CrossPlatform(Android), CrossPlatform(iOS), and CICIoT2022 datasets}
\label{tab:comparison_1}

\renewcommand{\arraystretch}{1.0} 
\resizebox{\textwidth}{!}{
\begin{tabular}{@{}lcccccccccccc@{}}
\toprule
\multirow{2}{*}{Method} & \multicolumn{4}{c}{CrossPlatform(Android)} & \multicolumn{4}{c}{CrossPlatform(iOS)} & \multicolumn{4}{c}{CICIoT2022} \\
\cmidrule(lr){2-5} \cmidrule(lr){6-9} \cmidrule(lr){10-13}
 & AC & PR & RC & F1 & AC & PR & RC & F1 & AC & PR & RC & F1 \\
\midrule
AppScanner    & 0.1626 & 0.1646 & 0.1456 & 0.1413 & 0.1718 & 0.1400 & 0.1440 & 0.1283 & 0.7556 & 0.8093 & 0.7244 & 0.6938 \\
FlowPrint     & 0.8739 & 0.8941 & 0.8739 & 0.8700 & 0.8712 & 0.8687 & 0.8712 & 0.8603 & 0.5820 & 0.4164 & 0.5820 & 0.4643 \\
FS-Net        & 0.0147 & 0.0023 & 0.0147 & 0.0034 & 0.0293 & 0.0014 & 0.0293 & 0.0025 & 0.5747 & 0.3800 & 0.5747 & 0.4216 \\
ET-BERT       & 0.9386 & 0.9451 & 0.9386 & 0.9401 & 0.9105 & 0.8809 & 0.9105 & 0.8850 & 0.9937 & 0.9938 & 0.9937 & 0.9937 \\
YaTC          & 0.9042 & 0.9081 & 0.9042 & 0.9042 & 0.9310 & 0.9307 & 0.9310 & 0.9295 & 0.9959 & 0.9959 & 0.9959 & 0.9959 \\
TrafficFormer & 0.7664 & 0.6435 & 0.6204 & 0.6167 & 0.5679 & 0.4966 & 0.4697 & 0.4689 & 0.8725 & 0.8487 & 0.8343 & 0.8288 \\
FlowletFormer & -      & -      & -      & -      & -      & -      & -      & -      & 0.9109 & 0.8905 & 0.8866 & 0.8859 \\
NetMamba      & 0.9869 & 0.9871 & 0.9869 & 0.9864 & 0.9881 & \textbf{0.9885} & 0.9881 & 0.9881 & 0.9985 & 0.9985 & 0.9985 & 0.9985 \\
MMAE       & \textbf{0.9897} & \textbf{0.9898} & \textbf{0.9897} & \textbf{0.9893} & \textbf{0.9900} & 0.9872 & \textbf{0.9900} & \textbf{0.9881} & \textbf{0.9990} & \textbf{0.9990} & \textbf{0.9990} & \textbf{0.9990} \\
\bottomrule
\end{tabular}
}
\end{table*}

\subsubsection{Implementation Details}

The teacher model and the student model have the same network architecture, which comprises an encoder consisting of 7 Transformer blocks and a decoder consisting of 2 Transformer blocks.

During the pre-training phase, we train the model for 150,000 iterations with a batch size of 128 using the AdamW optimizer. A linear learning rate scaling rule is applied with a base learning rate of $1 \times 10^{-3}$. For the teacher's EMA momentum parameters, $m_{base}$ and $m_{final}$ are set to 0.96 and 0.99, respectively. For the objective function, $\lambda_1$ and $\lambda_2$ are set to 1.0 and 0.1, respectively. Furthermore, the random masking ratio $r_{rand}$ is set to 0.7, supplemented by a dynamic hard masking ratio $r_{hard}$ of 0.2.

During the fine-tuning phase, the model is optimized over 120 epochs using the AdamW, with the base learning rate adjusted to $2 \times 10^{-3}$ and the batch size set to 64. Our framework is implemented in PyTorch 2.0.0 and trained on a single NVIDIA GeForce RTX 4080 GPU.

\subsubsection{Evaluation Metrics}
To evaluate the classification performance of our model, we employ four standard metrics: Accuracy (AC), Precision (PR), Recall (RC), and Weighted F1-Score (F1), formulated as:
\begin{equation}
\text{Weighted F1} = \sum_{i=1}^K \left( \frac{n_i}{N_{total}} \times \frac{2 \times \text{Precision}_i \times \text{Recall}_i}{\text{Precision}_i + \text{Recall}_i} \right)
\end{equation}
where $n_i$ is the number of samples in the $i$-th class, $N_{total}$ is the total number of samples, and $K$ is class number. 

\begin{table*}[htbp]
\centering
\caption{Comparison with previous SoTA methods on ISCXTor2016, ISCXVPN2016, and USTC-TFC2016 datasets}
\label{tab:comparison_2}
\renewcommand{\arraystretch}{1.0}
\resizebox{\textwidth}{!}{
\begin{tabular}{@{}lcccccccccccc@{}}
\toprule
\multirow{2}{*}{Method} & \multicolumn{4}{c}{ISCXTor2016} & \multicolumn{4}{c}{ISCXVPN2016} & \multicolumn{4}{c}{USTC-TFC2016} \\
\cmidrule(lr){2-5} \cmidrule(lr){6-9} \cmidrule(lr){10-13}
 & AC & PR & RC & F1 & AC & PR & RC & F1 & AC & PR & RC & F1 \\
\midrule
AppScanner    & 0.4034 & 0.2850 & 0.2149 & 0.2113 & 0.7643 & 0.8047 & 0.7045 & 0.7256 & 0.6998 & 0.8591 & 0.6062 & 0.6633 \\
FlowPrint     & 0.1316 & 0.0173 & 0.1316 & 0.0306 & 0.9666 & 0.9733 & 0.9666 & 0.9681 & 0.7992 & 0.7745 & 0.7992 & 0.7755 \\
FS-Net        & 0.7020 & 0.7010 & 0.7020 & 0.6999 & 0.7023 & 0.7487 & 0.7023 & 0.6660 & 0.4381 & 0.2011 & 0.4381 & 0.2672 \\
ET-BERT       & 0.9980 & 0.9981 & 0.9980 & 0.9980 & 0.9566 & 0.9566 & 0.9566 & 0.9565 & 0.9910 & 0.9911 & 0.9910 & 0.9910 \\
YaTC          & 0.9959 & 0.9959 & 0.9959 & 0.9959 & 0.9819 & 0.9820 & 0.9819 & 0.9819 & 0.9947 & 0.9749 & 0.9747 & 0.9734 \\
TrafficFormer & 0.8669 & 0.7545 & 0.7460 & 0.7472 & 0.8533 & 0.8445 & 0.8348 & 0.8279 & 0.9750 & 0.9789 & 0.9750 & 0.9746 \\
FlowletFormer & 0.9215 & 0.9263 & 0.9043 & 0.9116 & 0.9400 & 0.9471 & 0.9277 & 0.9364 & 0.9650 & 0.9689 & 0.9650 & 0.9648 \\
MLETC         & 0.9948 & 0.9778 & 0.9954 & 0.9865 & 0.9875 & 0.9847 & 0.9847 & 0.9880 & 0.9902 & 0.9935 & 0.9918 & 0.9926 \\
NetMamba      & 0.9993 & 0.9993 & 0.9993 & 0.9993 & 0.9899 & 0.9899 & 0.9899 & 0.9899 & 0.9990 & 0.9991 & 0.9990 & 0.9990 \\
MMAE       & \textbf{1.0000} & \textbf{1.0000} & \textbf{1.0000} & \textbf{1.0000} & \textbf{0.9957} & \textbf{0.9957} & \textbf{0.9957} & \textbf{0.9957} & \textbf{0.9995} & \textbf{0.9995} & \textbf{0.9995} & \textbf{0.9995} \\
\bottomrule
\end{tabular}
}
\end{table*}

\subsubsection{Comparison Methods}

We comprehensively evaluate and compare MMAE against diverse state-of-the-art methods, specifically including the following paradigms:

\begin{itemize}
    \item
    \textbf{Traditional Machine Learning:} AppScanner \cite{taylor2017robust} and FlowPrint \cite{Flowprint}, which rely on hand-crafted statistical features for classification.
    
    \item
    \textbf{CNN-based Models:} FS-Net \cite{Fs_net}, which performs supervised learning directly on the raw network packets.
    
    \item
    \textbf{Transformer-based Models:} ET-BERT \cite{Et_bert}, YaTC \cite{YATC}, TrafficFormer \cite{Trafficformer}, FlowletFormer \cite{FlowletFormer}, and MLETC \cite{MLETC}. These architectures capture latent representations by treating traffic as either 2D visual images or 1D language sequences before downstream fine-tuning.
    
    \item
    \textbf{Mamba-based Models:} NetMamba \cite{Netmamba}, a pure Mamba architecture that processes network traffic as block-wise 1D sequences.
\end{itemize} 

Aside from the supervised baselines (AppScanner, FlowPrint, and FS-Net), all methods utilize a pre-training paradigm. Regarding the pre-training corpora, TrafficFormer and FlowletFormer are pre-trained on the ISCX-VPN2016, CICIDS2017, and WIDE datasets \cite{kenjiro2000traffic}, whereas MLETC employs a proprietary dataset encompassing diverse real-world protocols.

\begin{table}[t]
\centering
\caption{Comparison with SoTA methods on the CSTNET-TLS dataset}
\label{tab:comparison_3}
\renewcommand{\arraystretch}{1.0} 
\begin{tabular}{@{}lcccc@{}}
\toprule
\multirow{2}{*}{Method} & \multicolumn{4}{c}{CSTNET-TLS} \\
\cmidrule(l){2-5}
 & AC & PR & RC & F1 \\
\midrule
AppScanner    & 0.7441 & 0.7232 & 0.6963 & 0.7023 \\
FS-Net        & 0.7814 & 0.7670 & 0.7316 & 0.7311 \\
ET-BERT       & 0.7993 & 0.7832 & 0.7689 & 0.7700 \\
YaTC          & 0.8391 & 0.8364 & 0.8101 & 0.8140 \\
TrafficFormer & 0.7982 & 0.7883 & 0.7736 & 0.7704 \\
FlowletFormer & 0.8605 & 0.8578 & 0.8445 & 0.8473 \\
NetMamba      & 0.9258 & 0.9271 & 0.9258 & 0.9249 \\
MMAE       & \textbf{0.9435} & \textbf{0.9459} & \textbf{0.9435} & \textbf{0.9434} \\
\bottomrule
\end{tabular}
\end{table}

\begin{table*}[htbp]
\centering
\caption{Ablation experiments of each component in MMAE}
\label{tab:ablation}
\renewcommand{\arraystretch}{1.0} 
\resizebox{\textwidth}{!}{
\begin{tabular}{@{}lcccccccccccc@{}}
\toprule
\multirow{2}{*}{Method} & \multicolumn{4}{c}{CrossPlatform(Android)} & \multicolumn{4}{c}{CrossPlatform(iOS)} & \multicolumn{4}{c}{CSTNET-TLS} \\
\cmidrule(lr){2-5} \cmidrule(lr){6-9} \cmidrule(lr){10-13}
 & AC & PR & RC & F1 & AC & PR & RC & F1 & AC & PR & RC & F1 \\
\midrule
Baseline     & 0.9852 & 0.9855 & 0.9852 & 0.9850 & 0.9838 & 0.9836 & 0.9836 & 0.9829 & 0.9271 & 0.9284 & 0.9271 & 0.9267 \\
+ SD & 0.9856 & 0.9858 & 0.9856 & 0.9854 & 0.9843 & 0.9855 & 0.9848 & 0.9840 & 0.9284 & 0.9302 & 0.9284 & 0.9280 \\
+ DMM        & 0.9878 & 0.9876 & 0.9878 & 0.9873 & 0.9881 & 0.9872 & 0.9881 & 0.9866 & 0.9336 & 0.9356 & 0.9336 & 0.9332 \\
+ SFM        & 0.9882 & 0.9882 & 0.9882 & 0.9880 & 0.9883 & 0.9857 & 0.9883 & 0.9866 & 0.9353 & 0.9361 & 0.9353 & 0.9346 \\
+ PMP       & \textbf{0.9897} & \textbf{0.9898} & \textbf{0.9897} & \textbf{0.9893} & \textbf{0.9900} & \textbf{0.9872} & \textbf{0.9900} & \textbf{0.9881} & \textbf{0.9435} & \textbf{0.9459} & \textbf{0.9435} & \textbf{0.9434} \\
\bottomrule
\end{tabular}
}
\end{table*}

\subsection{Main Results}

The comprehensive results are summarized in Tables~\ref{tab:comparison_1}, \ref{tab:comparison_2}, and \ref{tab:comparison_3}. MMAE consistently achieves state-of-the-art performance across all seven datasets, delivering F1-scores ranging from 0.9434 to 1.0000. While traditional machine learning and early deep learning methods capture primary statistical features, they exhibit pronounced limitations in complex scenarios. In contrast, MMAE sustains remarkable accuracy across all tasks. Although large Transformer-based architectures (e.g., ET-BERT, TrafficFormer, FlowletFormer, and MLETC) demonstrate baseline applicability, they suffer from adaptability and stability issues in traffic classification. For instance, TrafficFormer experiences severe performance degradation on cross-platform datasets (yielding F1-scores of merely 0.6167 and 0.4689) in Table~\ref{tab:comparison_1}. Similarly, image-based YaTC introduces unnecessary spatial interference. Standing apart from these NLP- or CV-inspired approaches, MMAE aligns intrinsically with the essence of network traffic.

\textbf{Performance on cross-platform and IoT traffic datasets}. MMAE achieves substantial improvements on cross-platform datasets. As shown in Table~\ref{tab:comparison_1}, MMAE reaches an accuracy of 0.9897 and an F1-score of 0.9893, delivering a significant 5.11\% accuracy improvement over ET-BERT on the CrossPlatform (Android) dataset. On the CrossPlatform (iOS) dataset, MMAE surpasses all baselines with an accuracy of 0.9900, outperforming YaTC by 5.90\%. Furthermore, on the IoT traffic dataset CICIoT2022, MMAE attains an accuracy of 0.9990.

\textbf{Performance on highly encrypted and malicious traffic datasets}. As shown in Table~\ref{tab:comparison_2}, MMAE demonstrates exceptional analytical efficacy. On the ISCXTor2016 dataset, all evaluation metrics achieve a perfect score of 1.0000. On the ISCXVPN2016 and USTC-TFC datasets, the accuracies reach 0.9957 and 0.9995, respectively. These results indicate that our designed architecture (a twin of MAE) and masking strategy (FlowMix) better capture complex dependencies within multi-layer encrypted packets, maintaining higher stability in encrypted environments.

\textbf{Performance under TLS 1.3 protocol}. Table~\ref{tab:comparison_3} highlights the limitations of existing models under the highly secure TLS 1.3 protocol. As shown in the Table~\ref{tab:comparison_3}, most baseline models witness significant performance degradation. In such restrictive scenarios, FS-Net leverages the packet-size sequence feature. 
By contrast, MMAE explicitly utilizes these key discriminant features by employing a multi-granularity modeling paradigm. As a result, it achieves an optimal accuracy of 0.9435 and an F1-score of 0.9434, outperforming the second-best method (NetMamba) by 1.77\% and 1.85\%, respectively. The overall capacity of MMAE is validated.

\subsection{Ablation Study}
To show the effect of each component in MMAE, we conduct an ablation study on the proposed method using a standard masked autoencoder with a random masking strategy as a baseline. We incrementally integrate Self-Distillation (+SD), Dynamic Mixed Masking (+DMM), Statistics-based Flow Matcher (+SFM), and the Packet-importance aware Mask Predictor (+PMP). Table~\ref{tab:ablation} details the results on the CrossPlatform (Android and iOS), which has the largest number of categories and the highly challenging CSTNET-TLS dataset.

Notably, the performance gains are most pronounced on CSTNET-TLS dataset, highlighting the superiority of the teacher-student based twin architecture in decoding complex modern encryption protocols. Specifically, adding the unmasked teacher (+SD) yields consistent improvements under multiple metrics, and introducing DMM brings a further performance improvement. For instance, on CSTNET-TLS, DMM boosts F1-score from 0.9280 to 0.9332. This confirms that breaking single-flow boundaries to construct cross-flow mixed pretext tasks effectively forces the encoder to extract discriminative features under severe interference. Subsequently, the addition of SFM provides a modest but crucial enhancement by ensuring that the paired flows constitute meaningful hard samples. Finally, equipped with PMP yields a much positive impact. On the heavily encrypted CSTNET-TLS dataset, PMP increases accuracy from 0.9353 to 0.9435. By dynamically allocating masking probabilities based on semantic density, PMP resolves the flaw of random masking, thereby maximizing the efficacy of the multi-granularity pre-training paradigm.

\begin{figure*}[t]
\centering
\subfloat[CrossPlatform(Android)]{
		\includegraphics[scale=0.28]{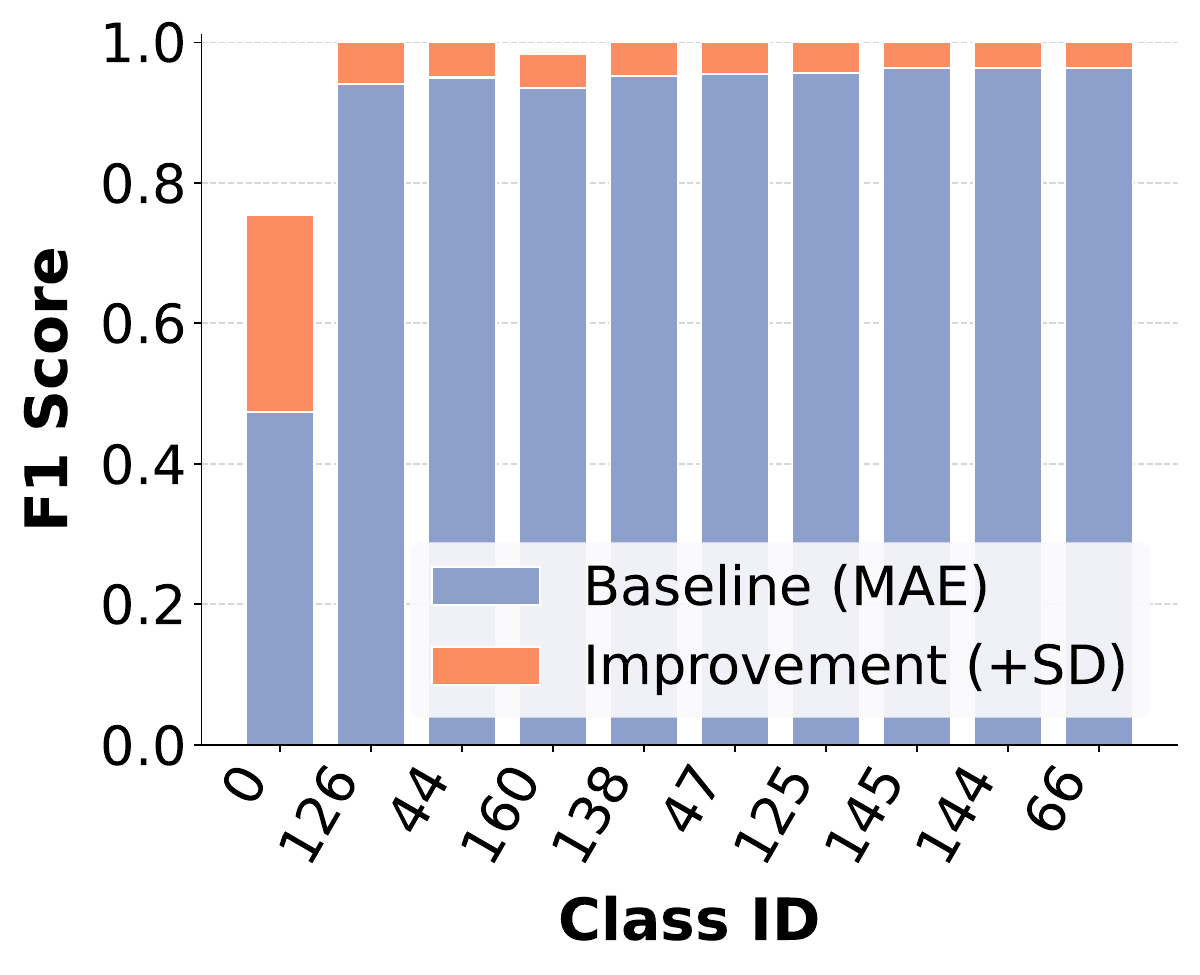}}
\subfloat[CrossPlatform(iOS)]{
		\includegraphics[scale=0.28]{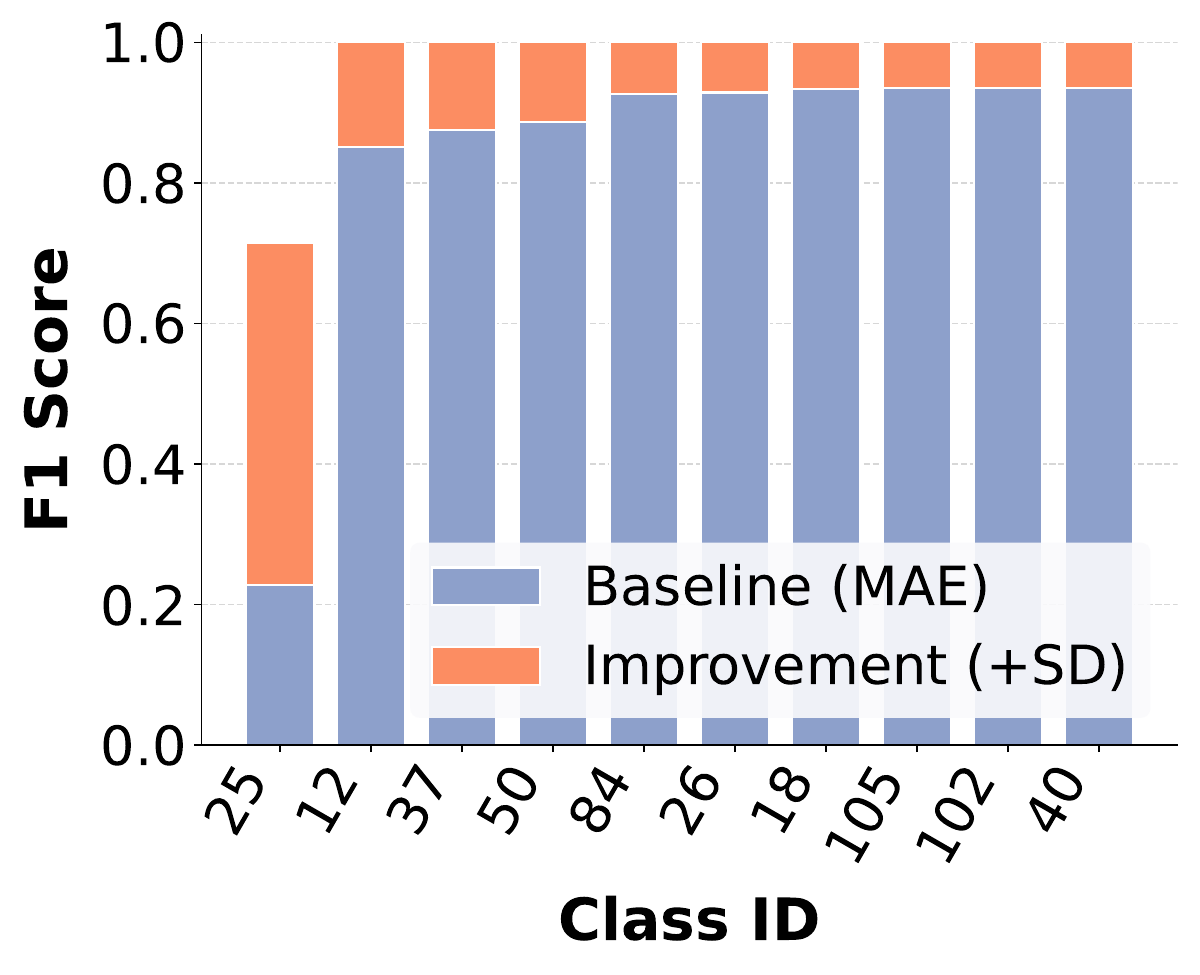}}
\subfloat[CSTNET-TLS 1.3]{
		\includegraphics[scale=0.28]{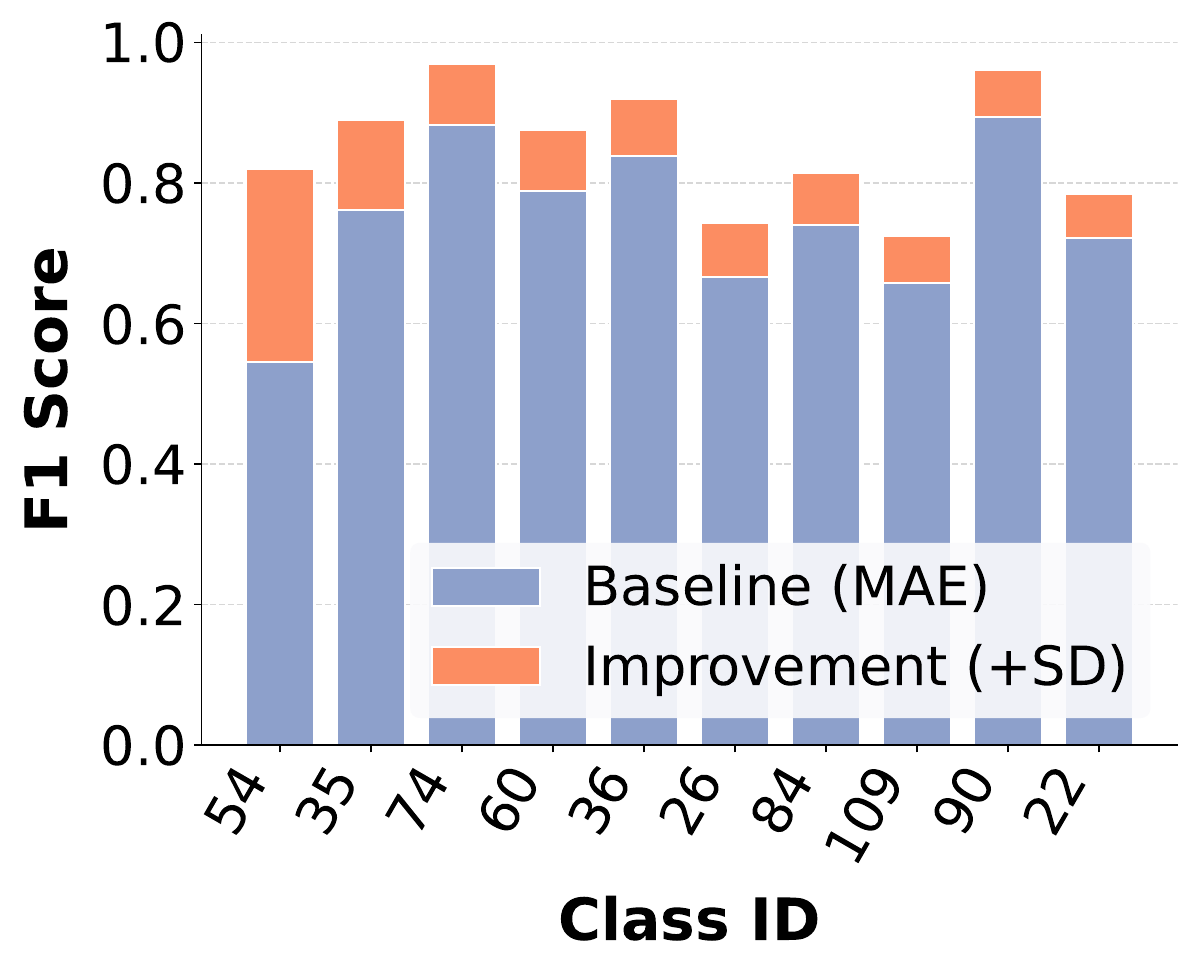}}
\caption{Comparison between MAE and MAE+SD on three datasets. The charts show 10 classes exhibiting prominent F1-score enhancements.
}
\label{F1_comparison}
\end{figure*}

\subsection{Visualization Results}

\textbf{F1-score visualization. }To demonstrate the effectiveness of our pre-training paradigm, Fig.~\ref{F1_comparison} compares the F1-scores of standard MAE against MAE with Self-Distillation (MAE+SD) across three datasets. 
We highlight 10 representative traffic classes with the most significant performance gains for each dataset. In the challenging CSTNET-TLS1.3 dataset, standard MAE struggles with many F1-scores falling below 0.80. Similarly, despite the overall better baseline performance, the two cross-platform datasets still suffer from severe bottleneck classes (e.g., Class 0 in Android and Class 25 in iOS). This highlights that relying solely on localized byte-level reconstruction is inadequate for capturing the discriminative semantics required for such complex traffic. By integrating self-distillation, the teacher branch provides flow-level global supervision, driving substantial enhancements across these difficult scenarios. For instance, the F1-score for Class 54 in CSTNET-TLS1.3 surges from 0.55 to 0.82. Meanwhile, the bottleneck classes in the cross-platform datasets have remarkable improvement. For instance, Class 25 in iOS is largely improved from 0.22 to over 0.70. These results comprehensively prove that multi-granularity semantic guidance effectively overcomes the limitations of standard MAE, significantly enhancing the model's discriminative capability.

\begin{figure*}[t]
 \centering
\subfloat[Pre-trained MAE]{
		\includegraphics[scale=0.14]{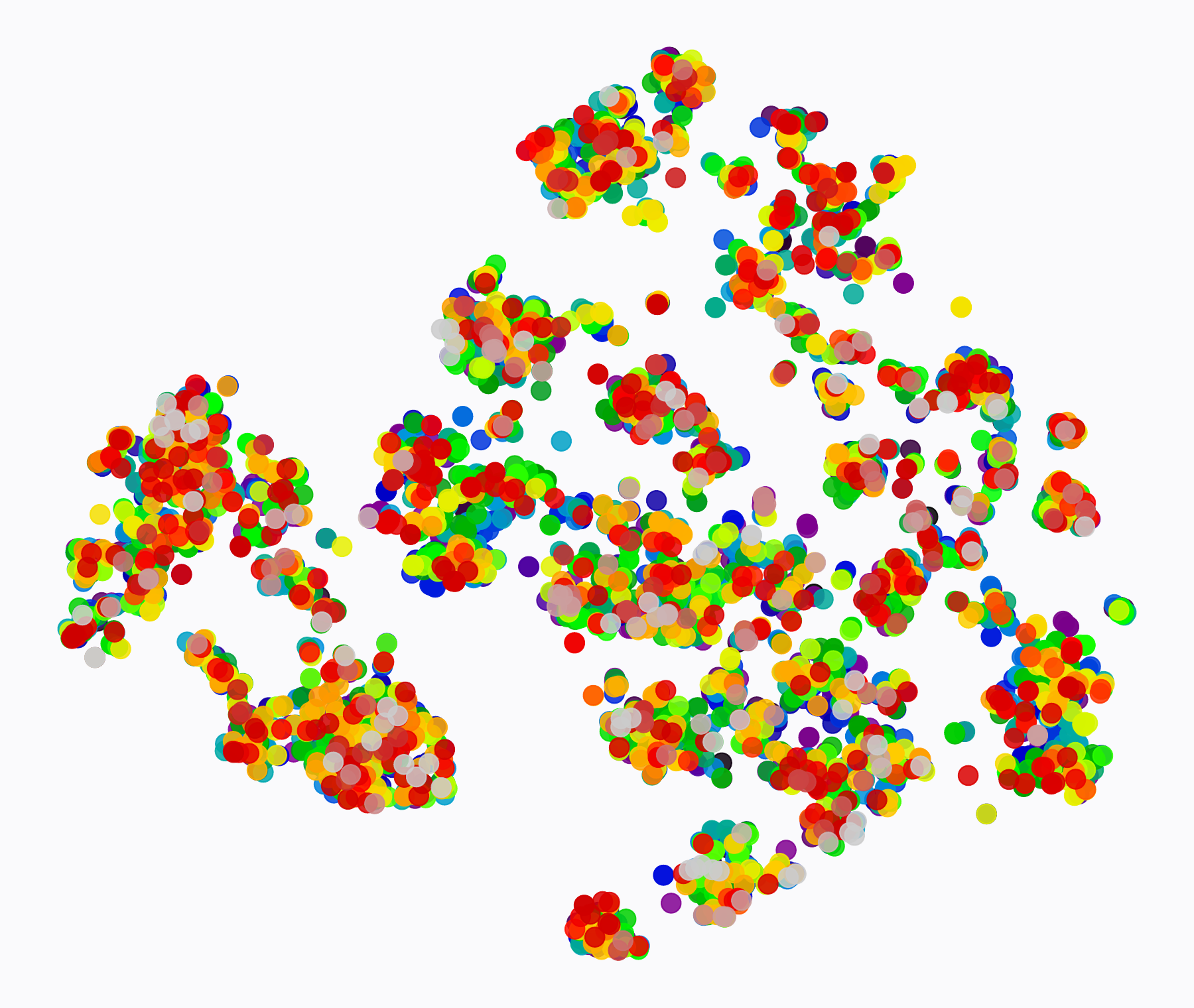}} \hspace{0.5mm}
\subfloat[Pre-trained MAE+SD]{
		\includegraphics[scale=0.14]{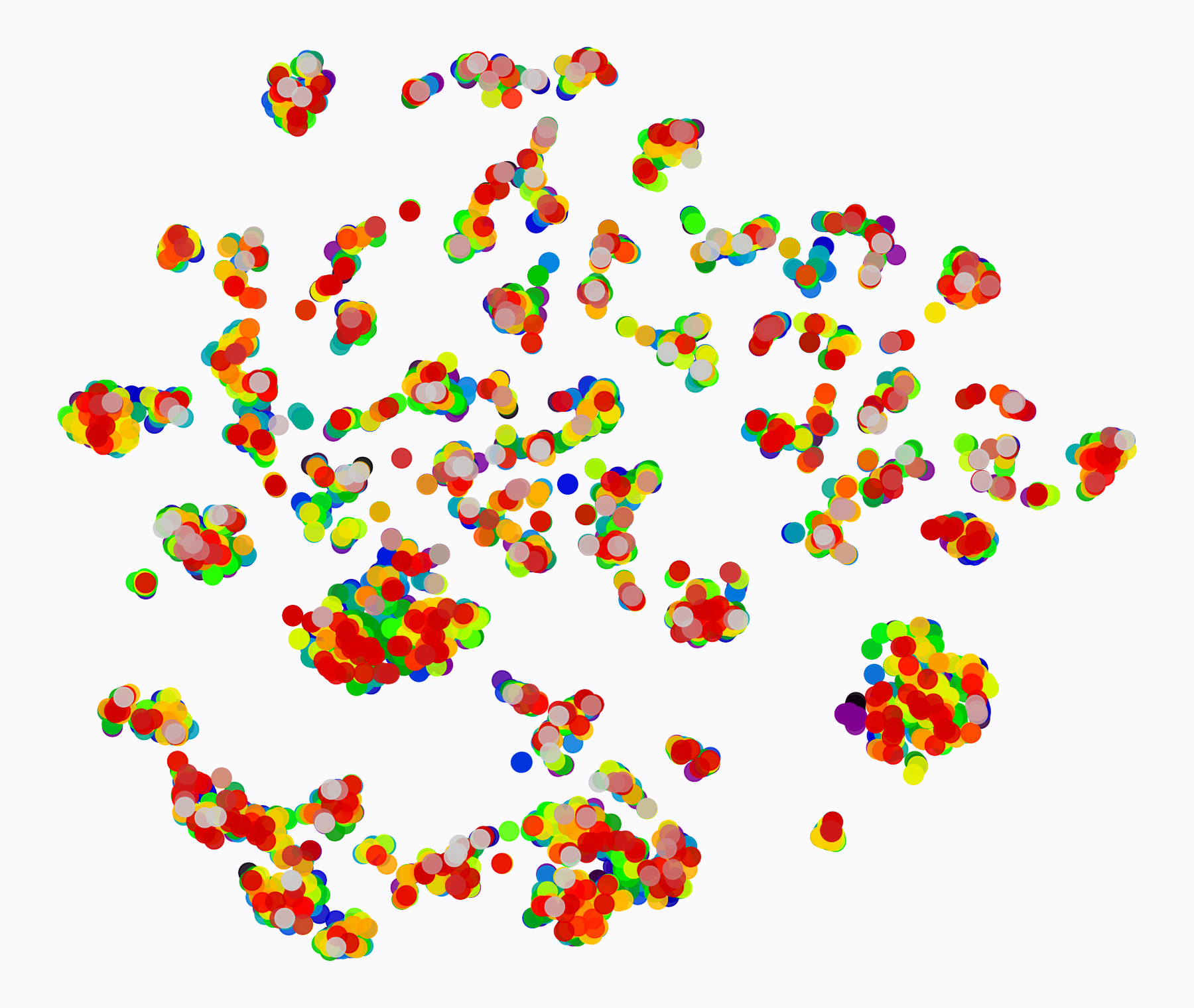}}  \hspace{0.5mm}
\subfloat[Fine-tuned MAE]{
		\includegraphics[scale=0.14]{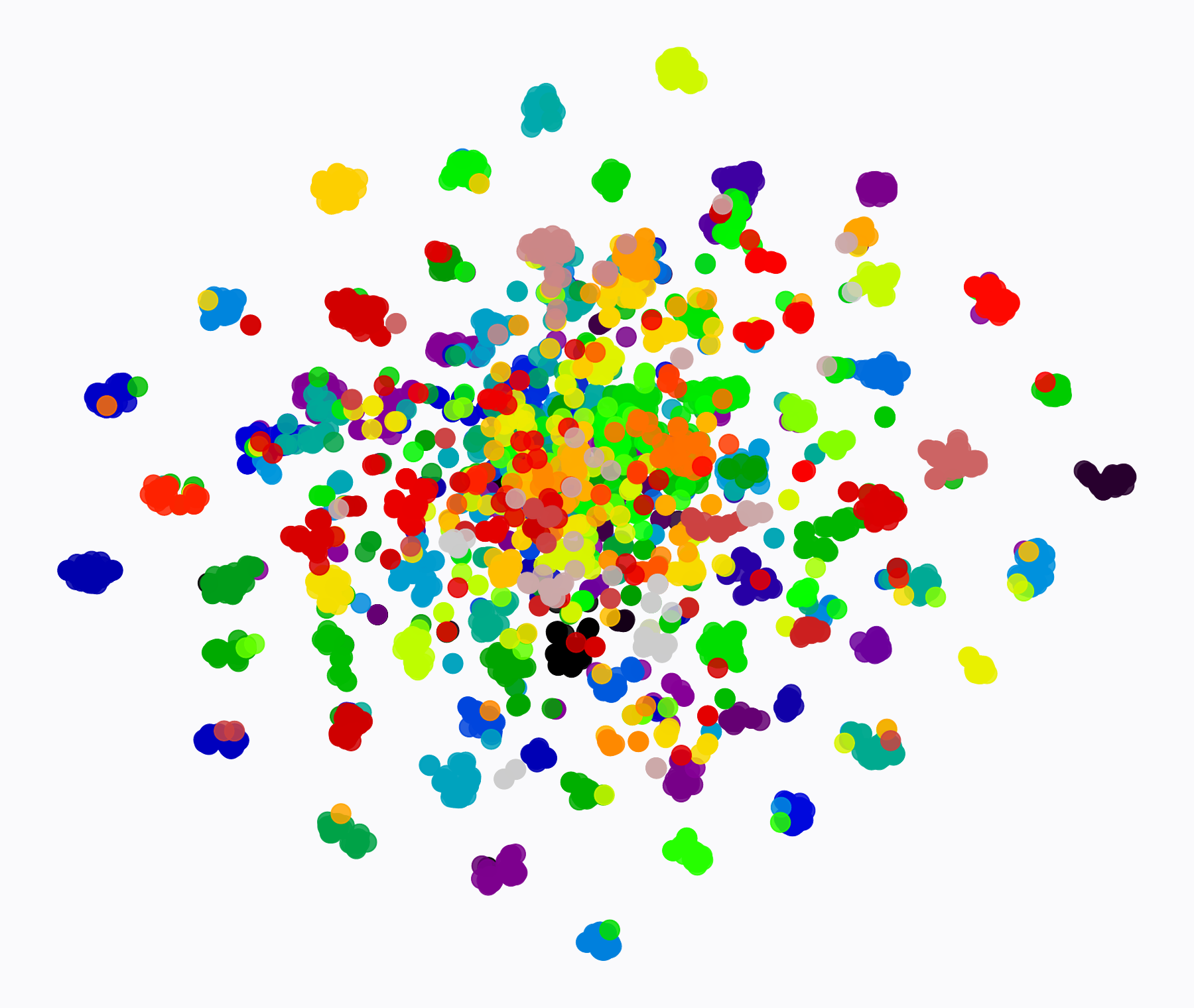}}  \hspace{0.5mm}
\subfloat[Fine-tuned MMAE (Ours)]{
		\includegraphics[scale=0.14]{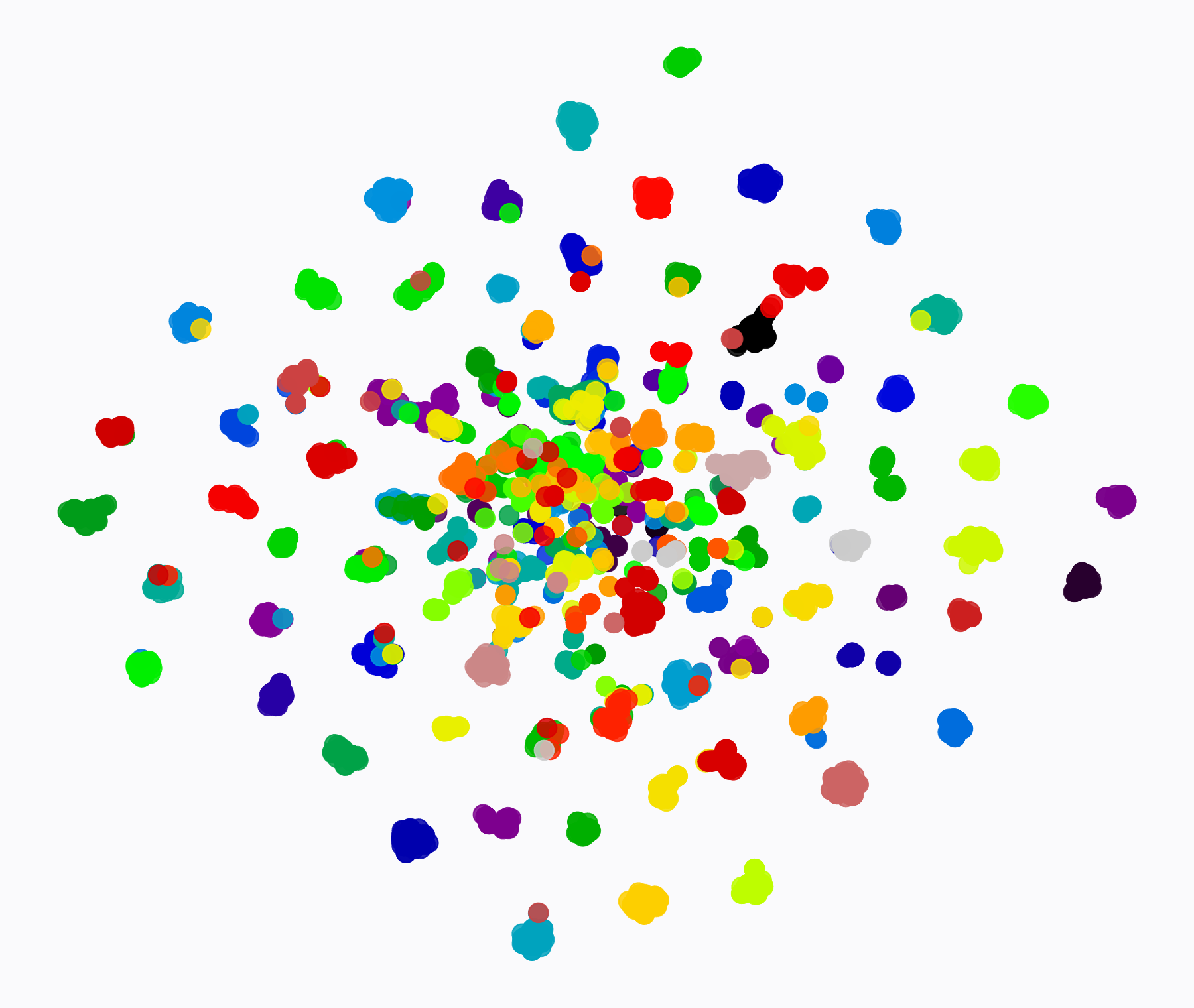}}  \hspace{0.5mm}
\caption{t-SNE visualization of hidden state representations across different training stages on the CSTNET-TLS 1.3 dataset. Distinct colors denote different traffic categories. MMAE effectively learns highly discriminative representations, exhibiting significantly clearer class boundaries compared to the baseline.
}

\label{t_sen}
\end{figure*}

\textbf{Visualization of feature distribution.} Fig.~\ref{t_sen} visualizes the t-SNE projections of the hidden state representations on the CSTNET-TLS 1.3 dataset, where different colors denote distinct categories. As observed, features from the pre-trained MAE are heavily entangled, while MAE+SD exhibits slightly better clustering. After fine-tuning via supervised cross-entropy loss, the separability of MAE in the feature space improves significantly. However, a large number of classes overlap in the central region. In contrast, the proposed MMAE yields highly discriminative representations, achieving distinguishable boundaries in the feature space.


\begin{figure*}[t]
 \centering
\subfloat[Throughput Comparison]{
		\includegraphics[scale=0.19]{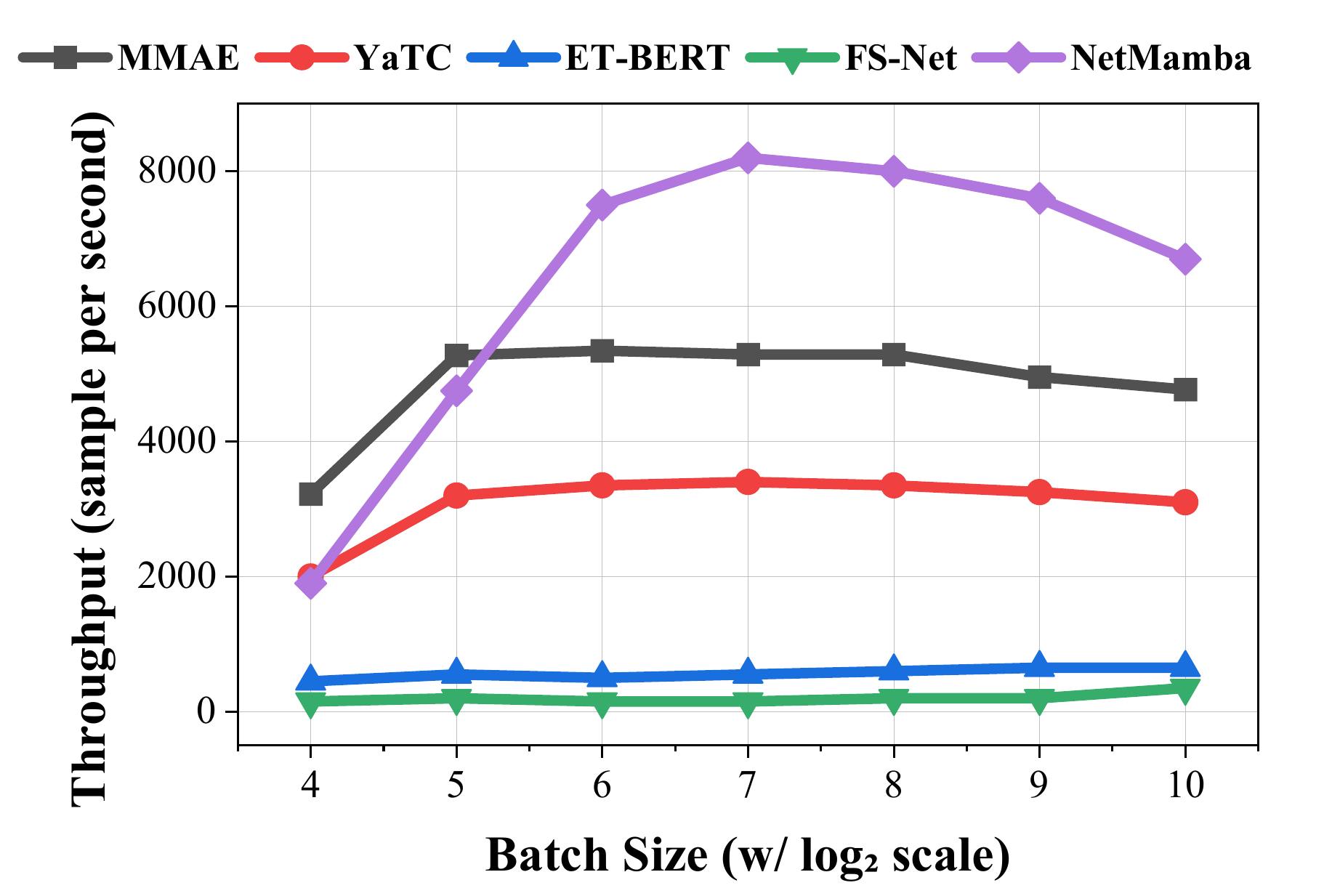}} \hspace{3mm}
\subfloat[Memory Comparison]{
		\includegraphics[scale=0.19]{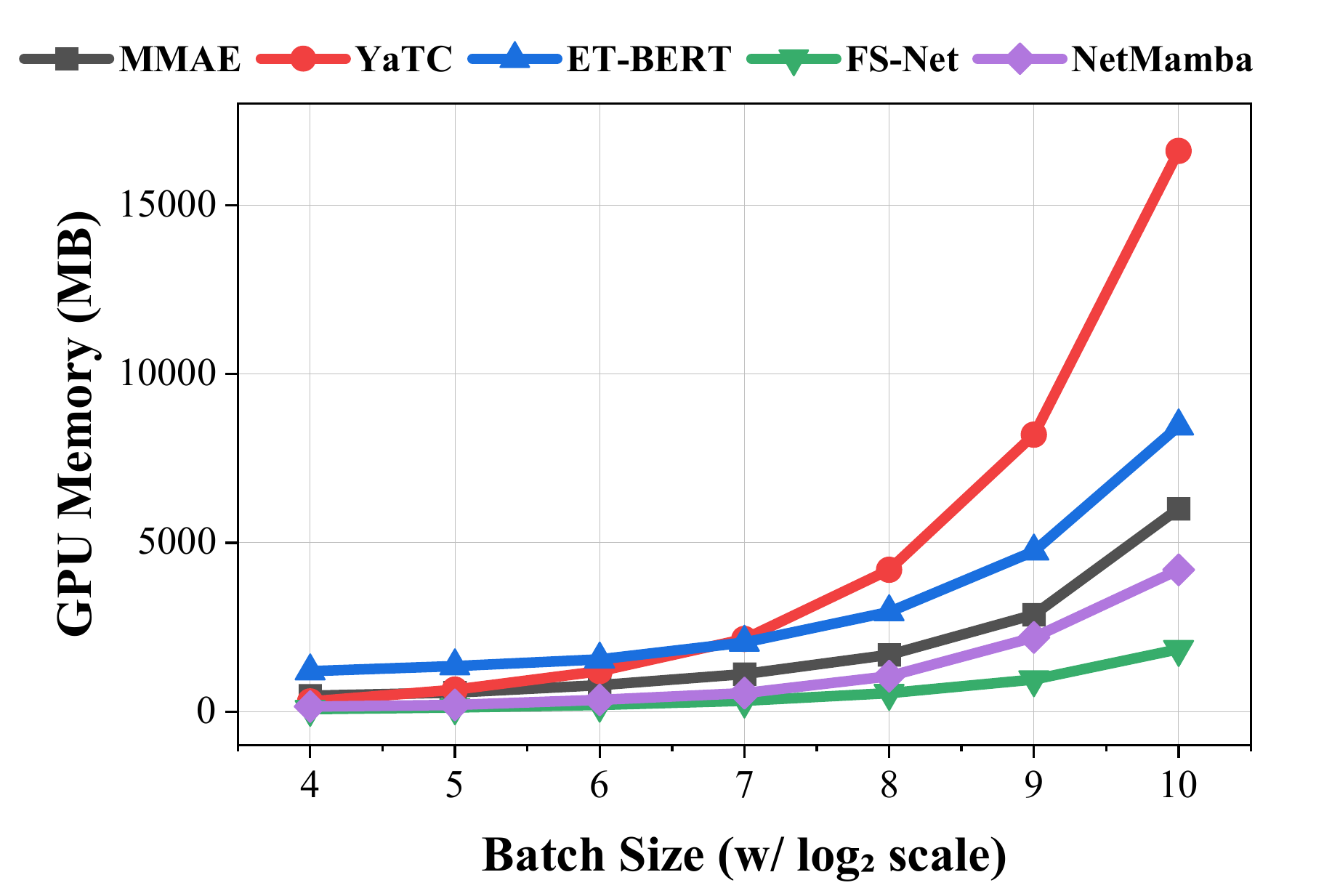}}  \hspace{1mm}
\subfloat[Inference Efficiency Comparison]{
		\includegraphics[width=0.29\linewidth]{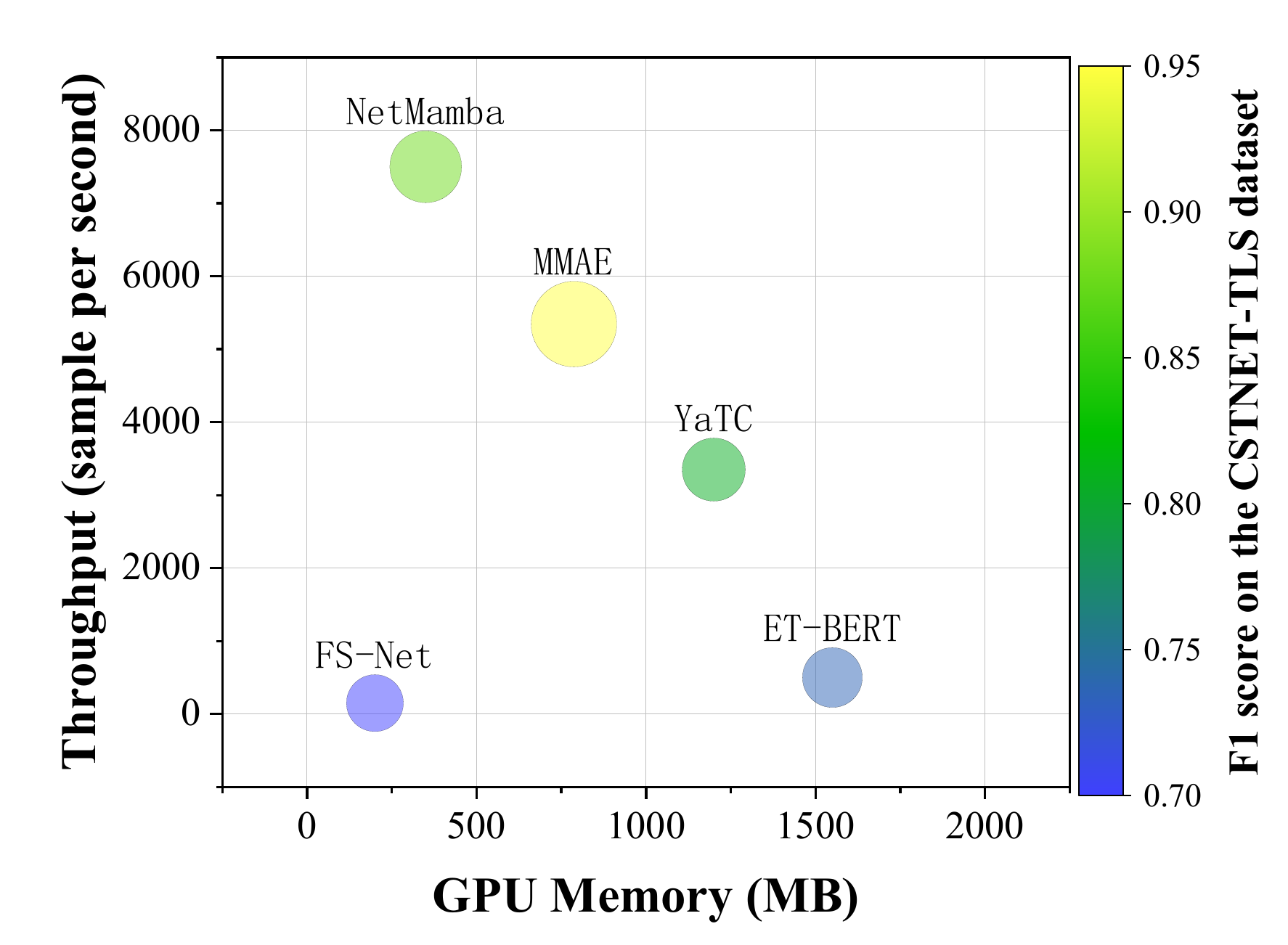}}
\caption{Efficiency evaluation of MMAE and baseline models. (a) Inference throughput, (b) GPU memory consumption across varying batch sizes, and (c) Comprehensive cost-effectiveness comparison at a batch size of 64.
}
\label{Efficiency}
\end{figure*}

\subsection{Computational Efficiency}

To evaluate the efficiency of MMAE, we compare its inference throughput and GPU memory consumption across batch sizes ranging from $2^4$ to $2^{10}$.

\textbf{Throughput comparison. }As illustrated in Fig.~\ref{Efficiency} (a), MMAE sustains a stable throughput of approximately 5,000 samples per second. It significantly outpaces large Transformer-based architectures, operating over $10\times$ faster than ET-BERT and nearly $1.6\times$ faster than YaTC. Although the pure state-space model NetMamba achieves the highest speed, MMAE remains highly competitive.

\textbf{Memory comparison. }Fig. \ref{Efficiency} (b) delineates the GPU memory footprint. MMAE demonstrates controlled memory growth, avoiding the steep memory overhead exhibited by YaTC and ET-BERT at larger batch sizes. Specifically, at a batch size of 64 ($2^6$), MMAE consumes approximately $50\%$ of the memory required by ET-BERT. While utilizing slightly more resources than the lightweight FS-Net and NetMamba, MMAE consumes substantially less memory than standard Transformer models.

\textbf{Inference efficiency comparison.} To illustrate the cost-effectiveness, Fig.~\ref{Efficiency} (c) presents a multi-dimensional comparison at a fixed batch size of 64. Ultimately, MMAE achieves the highest classification performance with moderate computational overhead.

\textbf{Efficiency and memory comparison of different phases.} Table~\ref{tab:efficiency} summarizes the computational overhead of MMAE across different phases. Despite its strong representational capabilities, the model maintains a lightweight parameter footprint and low GPU memory consumption.

\begin{table}[t]
\centering
\caption{Computational Efficiency Across Different Phases}
\label{tab:efficiency}
\resizebox{\columnwidth}{!}{%
\begin{tabular}{l c c l c c}
\toprule
\textbf{Phase} & \textbf{GPUs} & \textbf{Time} & \textbf{Unit/Granularity} & \textbf{Params} & \textbf{GPU Memory} \\
\midrule
Pre-training & 1 & 24.66 h & 5.918 s/100 steps & 8.65 M & 18GB \\
Fine-tuning  & 1 & 4.6 h   & 137.87 s / epoch  & 7.72 M & 3.6GB \\
Inference    & 1 & -       & 5343 samples/sec  & -      & 787MB \\
\bottomrule
\end{tabular}%
}
\end{table}

\subsection{Limitation Analysis}
PMP utilizes packet sizes and inter-arrival times to guide the masking process, which makes it susceptible to sophisticated adversarial traffic obfuscation. Advanced defense mechanisms may intentionally destroy these temporal and volumetric patterns, which may degrade PMP's masking efficacy. Furthermore, the pre-training phase incurs considerable computational overhead, although this is unavoidable for pre-training based paradigms. The optimization of teacher-student branches and the reliance on large batch sizes for cross-flow mixing require substantial GPU memory. Consequently, while inference remains lightweight, online pre-training on resource-constrained edge devices is currently impractical.

\section{Conclusion and Future Work}
In this paper, we propose MMAE, a teacher-student based twin MAE paradigm with
flow mixing strategy (FlowMix), to overcome the limitations of standard MAE in isolated byte-level reconstruction for encrypted traffic classification. By coupling a FlowMix strategy  with the PMP, our framework learns robust, multi-granularity semantic representations. Extensive evaluations confirm that MMAE establishes a new state-of-the-art pipeline across seven diverse datasets, demonstrating particular effectiveness in handling complex encryption protocols such as TLS 1.3.

Despite these promising results, several issues are still remained for future work. First, it is challenging to explore more robust feature extraction strategies that minimize susceptibility to adversarial obfuscation. Second, it is valuable to develop memory-efficient training pipeline to reduce computational overhead during the pre-training phase, enabling seamless deployment and online learning on resource-constrained edge devices. Finally, we intend to extend MMAE beyond standard classification to broader network management tasks, such as QoS prediction and malicious traffic detection.


%

\appendices


%
%

\ifCLASSOPTIONcaptionsoff
  \newpage
\fi



%

\bibliographystyle{IEEEtranS}
\bibliography{sn-bibliography}
\end{document}